\documentclass[sigconf]{acmart}

\usepackage{booktabs} 
\usepackage{flushend}
\usepackage[ruled]{algorithm2e} 
\usepackage{multirow, CJKutf8}
\usepackage{subfigure}
\usepackage{latexsym}
\usepackage{enumitem}
\usepackage{amsmath}
\usepackage{threeparttable}

\newcommand{\para}[1]{\smallskip\noindent{\bf {#1}. }}
\newcommand{\emoji}[1]{\includegraphics[width=1em]{emoji_images/#1.png}}
\usepackage{tikz}
\newcommand*{\circled}[1]{\lower.7ex\hbox{\tikz\draw (0pt, 0pt)%
    circle (.5em) node {\makebox[1em][c]{\small #1}};}}

\makeatletter
\def\subsubsection{\@startsection{subsubsection}{3}%
  \z@{.5\linespacing\@plus.7\linespacing}{.1\linespacing}%
  {\normalfont\itshape}}
\makeatother

\usepackage{pifont}
\usepackage{array}
\newcolumntype{L}[1]{>{\raggedright\let\newline\\\arraybackslash\hspace{0pt}}m{#1}}
\newcolumntype{C}[1]{>{\centering\let\newline\\\arraybackslash\hspace{0pt}}m{#1}}
\newcolumntype{R}[1]{>{\raggedleft\let\newline\\\arraybackslash\hspace{0pt}}m{#1}}

\newcommand{\approach}{ELSA\xspace}

\makeatletter
\def\subsubsection{\@startsection{subsubsection}{3}%
  \z@{.5\linespacing\@plus.7\linespacing}{.1\linespacing}%
  {\normalfont\itshape}}
\makeatother

\SetAlFnt{\small}
\SetAlCapFnt{\small}
\SetAlCapNameFnt{\small}
\SetAlCapHSkip{0pt}
\IncMargin{-\parindent}

\copyrightyear{2019}
\acmYear{2019} 
\setcopyright{iw3c2w3}
\acmConference[WWW '19]{Proceedings of the 2019 World Wide Web Conference}{May 13--17, 2019}{San Francisco, CA, USA}
\acmBooktitle{Proceedings of the 2019 World Wide Web Conference (WWW '19), May 13--17, 2019, San Francisco, CA, USA}
\acmPrice{}
\acmDOI{10.1145/3308558.3313600}
\acmISBN{978-1-4503-6674-8/19/05}

\begin{document}
\title{Emoji-Powered Representation Learning for Cross-Lingual Sentiment Classification}

\author{Zhenpeng Chen$^1$, Sheng Shen$^{1,2}$, Ziniu Hu$^{1,3}$, Xuan Lu$^1$, Qiaozhu Mei$^4$, Xuanzhe Liu$^1$}\authornote{Corresponding author: Xuanzhe Liu (xzl@pku.edu.cn).}
\affiliation{%
  \institution{$^1$Key Lab of High-Confidence Software Technology, MoE (Peking University), Beijing, China\\
  $^2$University of California, Berkeley, USA
  $^3$University of California, Los Angeles, USA\\
  $^4$School of Information, University of Michigan, Ann Arbor, USA
  }
 }
\email{czp@pku.edu.cn, sheng.s@berkeley.edu, bull@cs.ucla.edu, luxuan@pku.edu.cn, qmei@umich.edu, xzl@pku.edu.cn}

\begin{abstract}
Sentiment classification typically relies on a large amount of labeled data. In practice, the availability of labels is highly imbalanced among different languages, e.g., more English texts are labeled than texts in any other languages, which creates a considerable inequality in the quality of related information services received by users speaking different languages. To tackle this problem, cross-lingual sentiment classification approaches aim to transfer knowledge learned from one language that has abundant labeled examples (i.e., the source language, usually English) to another language with fewer labels (i.e., the target language). The source and the target languages are usually bridged through off-the-shelf machine translation tools. Through such a channel, cross-language sentiment patterns can be successfully learned from English and transferred into the target languages. This approach, however, often fails to capture sentiment knowledge specific to the target language, and thus compromises the accuracy of the downstream classification task. In this paper, we employ emojis, which are widely available in many languages, as a new channel to learn both the cross-language and the language-specific sentiment patterns. We propose a novel representation learning method that uses emoji prediction as an instrument to learn respective sentiment-aware representations for each language. The learned representations are then integrated to facilitate cross-lingual sentiment classification. The proposed method demonstrates state-of-the-art performance on benchmark datasets, which is sustained even when sentiment labels are scarce. 
\end{abstract}

%
%


%
%


\maketitle

\renewcommand{\shortauthors}{Z. Chen et al.}

\section{Introduction}\label{intro}
Sentiment analysis has become a critical topic in various research communities, including natural language processing (NLP)~\cite{DavidovTR10,FelboMSRL17}, Web mining~\cite{DeriuLLSMCHJ17,LiuZZHW18}, information retrieval~\cite{BalikasMA17,WuZYWHY17}, ubiquitous computing~\cite{Herdem12,SahaCBAC17}, and human-computer interaction~\cite{YangAAWL12,DiakopoulosS10}. Due to its effectiveness in understanding user attitudes, emotions, and even latent psychological statuses from text, sentiment analysis has been widely applied to all kinds of Web content such as blogs, Tweets, user reviews, and forum discussions, and it has been a critical component in many applications such as customer review tracking~\cite{Gamon04a}, sales prediction~\cite{LiuHAY07}, product ranking~\cite{McGlohonGR10}, stock market prediction~\cite{aclSiMLLLD13}, opinion polling~\cite{OConnorBRS10}, recommender systems~\cite{wwwSunGZ19}, personalized content delivery~\cite{HarakawaTOH18}, and online advertising~\cite{QiuHZSBC10}.

Similar to many other text mining tasks, existing work on sentiment analysis mainly deals with English texts~\cite{FelboMSRL17, LiuZZHW18,WuZYWHY17,DavidovTR10}. Although some efforts have also been made with other languages such as Japanese~\cite{PtaszynskiRAM12}, sentiment analysis for non-English languages is far behind. This creates a considerable inequality in the quality of the aforementioned Web services received by non-English users, especially considering that 74.6\% of Internet users are non-English speakers as of 2018~\cite{worlduser}. 
The cause of this inequality is quite simple: effective sentiment analysis tools are often built upon supervised learning techniques, and \textit{there are way more labeled examples in English than in other languages}. 

A straightforward solution is to transfer the knowledge learned from a label-rich language (i.e., the source language, usually English) to another language that has fewer labels (i.e., the target language), an approach known as \textit{cross-lingual sentiment classification}~\cite{ChenLL17}. In practice, the biggest challenge of cross-lingual sentiment classification is how to fill the linguistic gap between English and the target language. Many choose to bridge the gap through standard NLP techniques, and in particular, most recent studies have been using off-the-shelf machine translation tools to generate pseudo parallel corpora and then learn bilingual representations for the downstream sentiment classification task~\cite{PrettenhoferS10,XiaoG13, ZhouWX16}. More specifically, many of these methods enforce the aligned bilingual texts to share a unified embedding space, and sentiment analysis of the target language is conducted in that space. 

Although this approach looks sensible and easily executable, the performance of these machine translation-based methods often falls short. Indeed, a major obstacle of cross-lingual sentiment analysis is the so-called \textit{language discrepancy} problem~\cite{ChenLL17}, which machine translation does not tackle well. 
More specifically, sentiment expressions often differ a lot across languages. Machine translation is able to retain the general expressions of sentiments that are shared across languages (e.g., ``\textit{angry}'' or ``\begin{CJK}{UTF8}{min}怒っている\end{CJK}'' for negative sentiment), but it usually loses or even alters the sentiments in language-specific expressions~\cite{MohammadSK16}. As an example, in Japanese, the common expression ``\begin{CJK}{UTF8}{min}湯水のように使う\end{CJK}'' indicates a negative sentiment, describing the excessive usage or waste of a resource. However, its translation in English, ``\textit{use it like hot water},'' not only loses the negative sentiment but also sounds odd. 

The reason behind this pitfall is easy to explain: machine translation tools are usually trained on parallel corpora that are built in the first place to capture patterns shared across languages instead of patterns specific to individual languages.   
In other words, the problem is due to the failure to retain language-specific sentiment knowledge when unilaterally pursuing generalization across languages. A new bridge needs to be built beyond machine translation, which not only transfers ``general sentiment knowedge'' from the source language but also captures ``private sentiment knowledge'' of the target language. \textit{That bridge can be built with emojis}. 

In this paper, we tackle the problem of cross-lingual sentiment analysis by employing \textit{emojis} as an instrument. Emojis are considered an emerging ubiquitous language used worldwide~\cite{Lu16,BarbieriKRS16}; in our approach they serve both as a proxy of sentiment labels and as a bridge between languages.  
Their functionality of expressing emotions~\cite{HuGSNL17,CramerJT16} motivates us to employ emojis as complementary labels for sentiments, while their ubiquity~\cite{Lu16,BarbieriKRS16} makes it feasible to learn emoji-sentiment representations for almost every active language. 
Coupled with machine translation, the cross-language patterns of emoji usage can complement the pseudo parallel corpora and narrow the language gap, and the language-specific patterns of emoji usage help address the language discrepancy problem. 

We propose \approach, a novel framework of \textit{\underline{E}moji-powered representation learning for cross-\underline{L}ingual \underline{S}entiment \underline{A}nalysis}. Through \approach, language-specific representations are first derived based on modeling how emojis are used alongside words in each language. 
These per-language representations are then integrated and refined to predict the rich sentiment labels in the source language, through the help of machine translation. 
Different from the mandatorily aligned bilingual representations in existing studies, the joint representation learned through \approach catches not only the general sentiment patterns across languages, but also the language-specific patterns. In this way, the new representation and the downstream tasks are no longer dominated by the source language.


We evaluate the performance of \approach on a benchmark Amazon review dataset that has been used in various cross-lingual sentiment classification studies~\cite{PrettenhoferS10,XiaoG13,ZhouWX16}. The benchmark dataset covers nine tasks combined from three target languages (i.e., Japanese, French, and German) and three domains (i.e., book, DVD, and music). Results indicate that \approach outperforms existing approaches on all of these tasks in terms of classification accuracy. Experiments also show that the emoji-powered model works well even when the volume of unlabeled and labeled data are rather limited. To evaluate the generalizability of \approach, we also apply the method to Tweets, which again demonstrates state-of-the-art performance. In summary, the major contributions of this paper are as follows:

\begin{itemize}[leftmargin=*]
\item To the best of our knowledge, this is the first study that leverages emojis as an instrument in cross-lingual sentiment classification. We demonstrate that emojis provide not only surrogate sentiment labels but also an effective way to address language discrepancy. 

\item We propose a novel representation learning method to incorporating language-specific knowledge into cross-lingual sentiment classification, which uses an attention-based Long Short-Term Memory (LSTM) model to capture sentiments from emoji usage. 

\item We demonstrate the effectiveness and efficiency of \approach for cross-lingual sentiment classification using multiple large-scale datasets. \approach significantly improves the state-of-the-art results on the benchmark datasets.\footnote{The benchmark datasets, scripts, and pre-trained models are available at \url{https://github.com/sInceraSs/ELSA}.} 

\item The use of emojis as a bridge provides actionable insights into other Web mining applications that suffer from similar problem of inequality among languages. 

\end{itemize}

The rest of this paper is organized as follows. Section~\ref{related} presents the related work. Section~\ref{approach} formulates the problem and presents the proposed approach (\approach) to cross-lingual representation learning. Section~\ref{evaluation} evaluates \approach and analyzes the effectiveness of emojis in the learning process. Section~\ref{discussion} discusses the scalability and generalizability of \approach, followed by concluding remarks in Section~\ref{conclusion}.
\section{Related Work}\label{related}
We start with a summary of existing literature related to our study. 

\begin{figure*}
\begin{center}
\includegraphics[width=1.68\columnwidth]{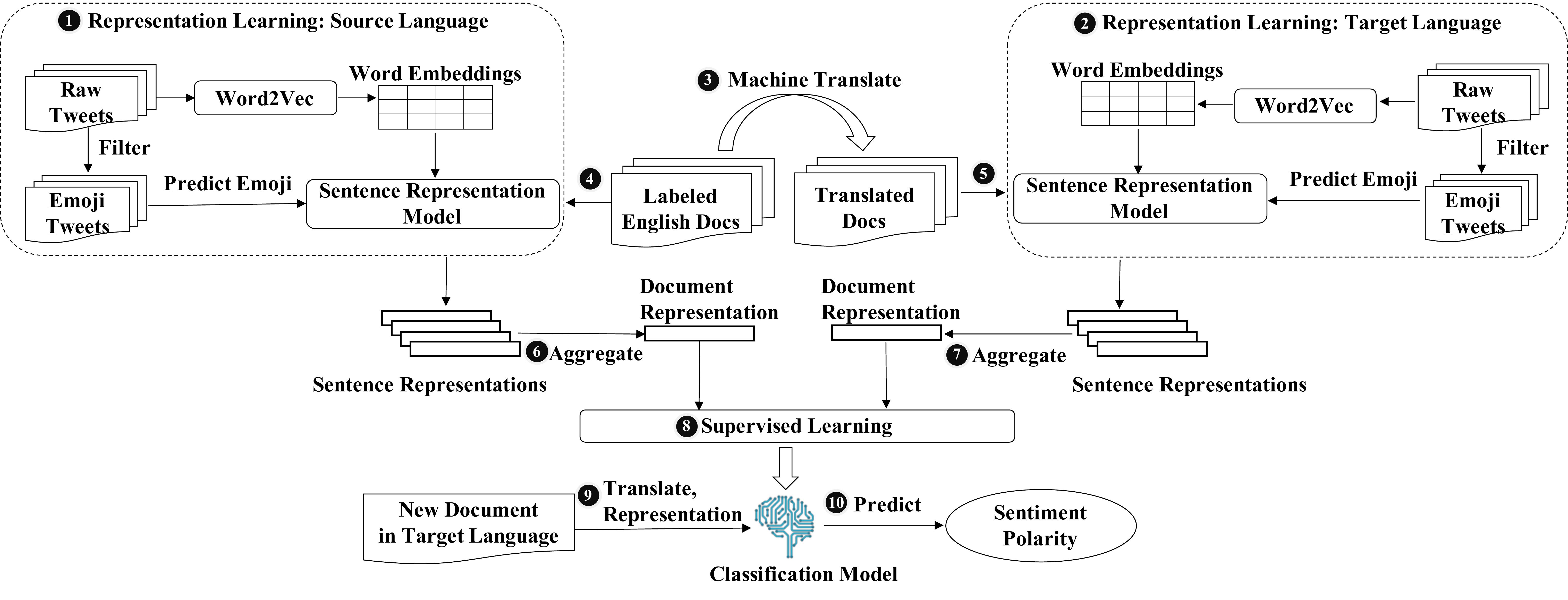}
\caption{The workflow of \approach.}\label{fig:approach}
\end{center}
\end{figure*}

\para{Emojis}
Emojis, also known as ideograms or smileys, can be used as compact expressions of objects, topics, and emotions. Being encoded in Unicode, they have no language barriers and are diffused on the Internet rapidly~\cite{Lu16}. The prevalence of emojis has attracted researchers from various research communities such as NLP, ubiquitous computing, human-computer interaction, multimedia, and Web mining~\cite{Lu16,CramerJT16,MillerTCJTH16,zhenpeng18,HuGSNL17,BarbieriKRS16,AiLLW0M17}. Many efforts have been devoted to studying their usage across platforms~\cite{MillerTCJTH16}, across genders~\cite{zhenpeng18}, across languages~\cite{BarbieriKRS16}, and across cultures~\cite{Lu16}. The various non-verbal functions of emojis play an important role in their wide adoption. Emojis are used to replace content words, express situational and additional emotions, adjust tones, express intimacy, etc.~\cite{HuGSNL17,CramerJT16}. In particular, expressing sentiment is demonstrated to be the most popular intention for using emojis~\cite{HuGSNL17}, so that emojis can be used as effective proxies for sentiment polarities~\cite{FelboMSRL17}. Considering the ubiquitous usage of emojis across languages and their functionality of expressing sentiments, we make the first effort to use emojis as an instrument to improve cross-lingual sentiment analysis.

\para{Textual Sentiment Analysis}
Sentiment analysis is a classical NLP task aiming to study the emotions, opinions, evaluations, appraisals, and attitudes of people from text data~\cite{2012Liu}. Many widely used tools, such as SentiStrength~\cite{ThelwallBPCK10} and LIWC~\cite{Pennebaker1999Linguistic}, simply aggregate the polarity of individual words to determine the overall sentiment score of a text. Better performance of sentiment classification is often obtained through supervised machine learning~\cite{pang2002thumbs}. 
Recently, with the emergence of deep learning techniques, many researchers have attempted to use advanced neural network models for sentiment analysis~\cite{ZhangWL18}. Supervised machine learning methods, including deep learning models, usually require a large volume of labeled data for training. In reality, however, high-quality sentiment labels are often scarce due to the labor-consuming and error-prone human annotation process~\cite{FelboMSRL17}. To address this limitation, researchers have used sentimental hashtags and emoticons as weak sentiment labels~\cite{DavidovTR10,DeriuLLSMCHJ17}. These weak labels are usually language/community-specific. In addition, figuring out the sentiment polarities of certain hashtags or emoticons can be hard. In recent years, emoticons have been gradually replaced by increasingly popular emojis~\cite{PavalanathanE15a}, and emojis have started to be explored as proxies of sentiment labels~\cite{FelboMSRL17}. We follow the same intuition and utilize emojis as surrogate labels to learn per-language representations. Instead of attempting to directly map emojis to sentiment polarities, however, we integrate these language-specific representations and feed them through downstream tasks to predict real, high quality sentiment labels (in the source language). 

\para{Cross-Lingual Text Classification}
There is a significant imbalance in the availability of labeled corpora among different languages: more in English, and much fewer in other languages. Cross-lingual learning is a common approach to tackling this problem in various text mining tasks such as Web page classification~\cite{LingXDJYY08}, topic categorization~\cite{ZhouPTH16}, and sentiment analysis~\cite{PrettenhoferS10,XiaoG13,ZhouWX16}. 
Many researchers divide cross-lingual learning process into two stages: first encoding texts in the source and the target languages into continuous representations, and then utilizing these representations for the final classification task in the target language~\cite{PrettenhoferS10,XiaoG13,ZhouWX16}. To bridge the linguistic gap between the source and the target languages, most studies introduce a translation oracle to project different languages' representations into a unified space at different (e.g., word or document) levels~\cite{PrettenhoferS10,PLLKRRS14,ZhouWX16, XiaoG13}. The performance of these methods thus heavily depends on the quality of the machine translation tools and the pseudo parallel corpora they generate. 
Unfortunately, different from topical words, emotional language patterns like sentiment (or sarcasm, humor), which present strong language-specific characteristics, cannot be easily transferred in this way.
We utilize the easily accessible emoji-texts to incorporate both cross-language and language-specific knowledge into the representations of the source and the target languages. The implicit sentiment knowledge encoded in the usage of diverse emojis solves both the label imbalance and the language discrepancy problems. 

\section{The \approach Approach}\label{approach}
To better illustrate the workflow of \approach, we first give a formulation of our problem. Cross-lingual sentiment classification aims to use the labeled data in a source language (i.e., English) to learn a model that can classify the sentiment of test data in a target language. In our setting, besides labeled English documents ($L_S$), we also have large-scale unlabeled data in English ($U_S$) and in the target language ($U_T$). Furthermore, there exist unlabeled data containing emojis, both in English ($E_S$) and in the target language ($E_T$). In practice, these unlabeled, emoji-rich data can be easily obtained from online social media such as Twitter. 
Our task is to build a model that can classify the sentiment polarity of document in the target language solely based on the labeled data in the source language (i.e., $L_S$) and the different kinds of unlabeled data (i.e., $U_S$, $U_T$, $E_S$ and $E_T$). Finally, we use a held-out set of labeled documents in the target language ($L_T$), which can be small, to evaluate the model.

The workflow of \approach is illustrated in Figure~\ref{fig:approach}, with the following steps. In \textit{step 1} and \textit{step 2}, we build sentence representation models for both the source and the target languages.  Specifically, for each language, we employ a large number of Tweets to learn word embeddings (through Word2Vec~\cite{Mikolov2013Efficient}) in an unsupervised fashion. From these word embeddings, we learn higher-level sentence representation through predicting the emojis used in a sentence. This can be viewed as a distantly supervised learning process, where emojis serve as surrogate sentiment labels. In \textit{step 3}, we translate each labeled English document into the target language, sentence by sentence, through \textit{Google Translate}. Both the English sentences and their translations are fed into the representation models learned in steps 1 and 2 to obtain their per-language representations (\textit{step 4} and \textit{step 5}). Then in \textit{step 6} and \textit{step 7} we aggregate these sentence representations back to form two compact representations for each training document, one in English and the other in the target language. In \textit{step 8}, we use the two representations as features to predict the real sentiment label of each document and obtain the final sentiment classifier. In the test phase, for a new document in the target language, we translate it into English and then follow the previous steps to obtain its representation (\textit{step 9}), based on which we predict the sentiment label using the classifier (\textit{step 10}). 

\subsection{Representation Learning}
Representations of documents need to be learned before we train the sentiment classifier. Intuitively, one could simply use off-the-shelf word embedding techniques to create word representations and then average the word vectors to obtain document embeddings. 
Such embeddings, however, capture neither per-language nor cross-language sentiment patterns. Since emojis are widely used to express sentiments across languages, we learn sentiment-aware representations of documents using emoji prediction as an instrument. 
Specifically, in a distantly supervised way, we use emojis as surrogate sentiment labels and learn sentence embeddings by predicting which emojis are used in a sentence. This representation learning process is conducted separately in the source and the target languages to capture language-specific sentiment expressions.

\begin{figure}[!tp]
\includegraphics[width=0.76\columnwidth]{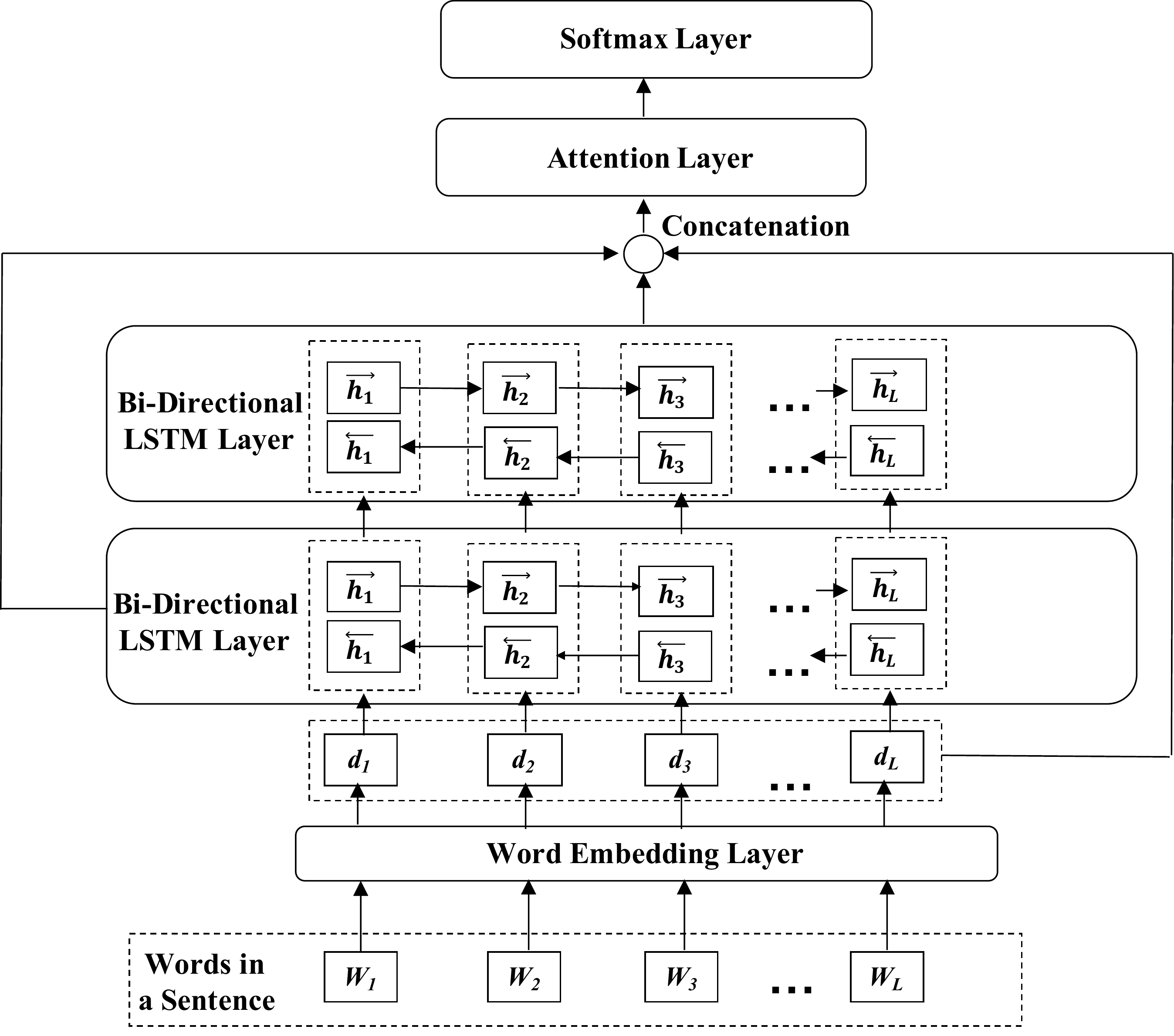}
\caption{Network architecture for representation learning through emoji prediction.}\label{fig:emoji_predict}
\end{figure}

The architecture of the representation learning model is illustrated in Figure~\ref{fig:emoji_predict}. First, we pre-train low-level word embeddings using tens of millions of unlabeled Tweets (i.e., the word embedding layer). Then, we represent every single word as a unique vector and use stacked bi-directional LSTM layers and one attention layer to encode these word vectors into sentence representations. The attention layer takes the outputs of both the embedding layer and the two LSTM layers as input, through the skip-connection algorithm~\cite{Hermans2013Training}, which enables unimpeded information flow in the whole training process. Finally, the model parameters are learned by minimizing the output error of the softmax layer. The details of the architecture are elaborated below.

\para{Word Embedding Layer}
The word embeddings are pre-trained with the skip-gram algorithm~\cite{Mikolov2013Efficient} based on either $U_S$ or $U_T$, which encode every single word into a continuous vector space. Words that commonly occur in a similar context are embedded closely in the vector space, which captures word semantic information. We leave the details of this standard Word2Vec process to the readers~\cite{Mikolov2013Efficient}. 

\para{Bi-Directional LSTM Layer} 
As a special type of recurrent neural network (RNN), LSTM~\cite{Hochreiter1997Long} is particularly suitable for modeling the sequential property of text data. 
At each step (e.g., word token), LSTM combines the current input and knowledge from the previous steps to update the states of the hidden layer. To tackle the gradient vanishing problem \cite{Hochreiter98} of traditional RNNs, LSTM incorporates a gating mechanism to determine when and how the states of hidden layers can be updated. Each LSTM unit contains a memory cell and three gates (i.e., an input gate, a forget gate, and an output gate)~\cite{PascanuMB13}. The input and output gates control the input activations into the memory cell and the output flow of cell activations into the rest of the network, respectively. The memory cells in LSTM store the sequential states of the network, and each memory cell has a self-loop whose weight is controlled by the forget gate. 

Let us denote each sentence in $E_S$ or $E_T$ as ($x$, $e$), where $x=[d_1,d_2,...,d_L]$ as a sequence of \textit{word} vectors representing the plain text (by removing emojis) and $e$ as one emoji contained in the text. At step $t$, LSTM computes unit states of the network as follows:




\begin{subequations}
\begin{equation*}
i^{(t)} = \sigma(U_id_t+W_ih^{(t-1)}+b_i),
\end{equation*}
\begin{equation*}
f^{(t)}=\sigma(U_fd_t+W_fh^{(t-1)}+b_f),
\end{equation*}
\begin{equation*}
o^{(t)}=\sigma(U_od_t+W_oh^{(t-1)}+b_o),
\end{equation*}
\begin{equation*}
c^{(t)}=f_t \odot c^{(t-1)} + i^{(t)} \odot tanh(U_cd_t+W_ch^{(t-1)}+b_c),
\end{equation*}
\begin{equation*}
h^{(t)} = o^{(t)} \odot tanh(c^{(t)}),
\end{equation*}
\end{subequations}
where $i^{(t)}$, $f^{(t)}$, $o^{(t)}$, $c^{(t)}$, and $h^{(t)}$ denote the state of the input gate, forget gate, output gate, memory cell, and hidden layer at step $t$. $W$, $U$, $b$ respectively denote the recurrent weights, input weights, and biases. $\odot$ is the element-wise product. We can extract the latent vector for each step $t$ from LSTM. In order to capture the information from the context both preceding and following a word, we use the bi-directional LSTM. We concatenate the latent vectors from both directions to construct a bi-directional encoded vector $h_i$ for every single word vector $d_i$, which is: 

\begin{subequations}
\begin{equation*}
\mathop{h_i}\limits ^{\rightarrow}=\mathop{LSTM}\limits ^{\longrightarrow}(d_i), i \in [1,L],
\end{equation*}
\begin{equation*}
\mathop{h_i}\limits ^{\leftarrow}=\mathop{LSTM}\limits ^{\longleftarrow}(d_i), i \in [L,1],
\end{equation*}
\begin{equation*}
h_i = [\mathop{h_i}\limits ^{\rightarrow}, \mathop{h_i}\limits ^{\leftarrow}].
\end{equation*}
\end{subequations}


\para{Attention Layer}
We employ a skip-connection that concatenates the outputs of the embedding layer and the two bi-directional LSTM layers as the input of the attention layer. The $i$-th word of the input sentence can be represented as $u_i$:

\begin{equation*}
u_i = [d_i, h_{i1}, h_{i2}],
\end{equation*}
where $d_i$, $h_{i1}$, and $h_{i2}$ denote the encoded vectors of words extracted in the word embedding layer and the first and second bi-directional LSTMs, respectively. Since not all words contribute equally to predicting emojis or expressing sentiments, we employ the attention mechanism~\cite{BahdanauCB14} to determine the importance of every single word. The attention score of the $i$-th word is calculated by

\begin{equation*}
a_i = \frac{exp(W_au_i)}{\sum_{j=1}^Lexp(W_au_j)},
\end{equation*}
where $W_a$ is the weight matrix used by the attention layer. Then each sentence can be represented as the weighted sum of all words in it, using the attention scores as weights. That is,

\begin{equation*}
v=\sum_{i=1}^La_iu_i.
\end{equation*}

\para{Softmax Layer}
The sentence representation is then transferred into the softmax layer, which returns a probability vector $Y$. Each element of this vector indicates the probability that this sentence contains a specific emoji. The $i$-th element of the probability vector is calculated as:

\begin{equation*}
y_i=\frac{exp(v^Tw_i+b_i)}{{\sum_{j=1}^K}exp(v^Tw_j+b_j)},
\end{equation*}
where $w_i$ and $b_i$ define the weight and bias of the $i$-th element. Finally, we learn the model parameters by minimizing the cross entropy between the output probability vectors and the one-hot vectors of the emoji contained in each sentence. After learning the parameters, we can extract the output of the attention layer to represent each input sentence. Through this emoji-prediction process, words with distinctive sentiments can be identified, and the plain text surrounding the same emojis will be represented similarly. 
Given the fact that the sentiment labels are limited, once the emoji-powered sentence representations are trained, they are locked in the downstream sentiment prediction task to avoid over-fitting. 

\subsection{Training the Sentiment Classifier}
Based on the pre-trained, per-language sentence representations, we then learn document representations and conduct cross-lingual sentiment classification. 

First, for each English document $D_s \in L_S$, we use the pre-trained English representation model to embed every single sentence in it. Second, we aggregate these sentence representations to derive a compact document representation. Because different parts of a document contribute differently to the overall sentiment, we once again adopt the attention mechanism here. Supposing the sentence vectors as $v_i$, we calculate the document vector $r_s$ as:
\begin{eqnarray*}
r_s = {\sum}_{i=1}^N{\beta}_iv_i, ~~~ \text{where}\\
{\beta}_i = \frac{exp(W_bv_i)}{\sum_{j=1}^Nexp(W_bv_j)},	
\end{eqnarray*}
where $W_b$ is the weight matrix of the attention layer and ${\beta}_i$ is the attention score of the $i$-th sentence in the document. Next, we use \textit{Google Translate} to translate $D_s$ into the target language ($D_t$). We then leverage the pre-trained target-language representation model to form representations for each translated document following the same process above. Supposing the text representations of $D_s$ and $D_t$ are $r_s$ and $r_t$ respectively, we concatenate them into a joint representation $r_c = [r_s,r_t]$, which contains sentiment knowledge from both English and the target language, ensuring that our model is not dominated by the labeled English documents. Finally, we input $r_c$ into an additional softmax layer to predict the real sentiment label of $D_s$. 

\subsection{Sentiment Classification for Target Language}
When we receive an unlabeled document in $L_T$, we first translate it into English. Based on the representation models trained above, the original document and its English translation can be represented as $r_t$ and $r_s$. We represent this document as [$r_s$, $r_t$] and input it into the classifier, which outputs a predicted sentiment polarity.


\section{Evaluation}\label{evaluation}
In this section, we evaluate the effectiveness and efficiency of \approach using standard benchmark datasets for cross-lingual sentiment classification as well as a large-scale corpus of Tweets.

\begin{table}[!tp]
\centering
\small
\caption{The sizes of the Tweets and emoji-Tweets.}
\label{data_size}
\begin{tabular}{lrrrr}
\hline
Language & English & Japanese & French & German \\
\hline
Raw Tweets& 39.4M & 19.5M  & 29.2M & 12.4M \\
\hline
Emoji-Tweets & 6.6M & 2.9M & 4.4M & 2.7M \\
\hline
\end{tabular}
\end{table} 

\subsection{The Dataset}
The labeled data ($L_S$ for training and $L_T$ for testing) used in our work are from the Amazon review dataset~\cite{Webis} created by Prettenhofer and Stein~\cite{PrettenhoferS10}. This dataset is representative and used in a variety of cross-lingual sentiment classification work~\cite{PrettenhoferS10,XiaoG13,ZhouWX16}. It covers four languages (i.e., English, Japanese, French, and German) and three domains (i.e., book, DVD, and music). For each combination of language and domain, the dataset contains 1,000 positive reviews and 1,000 negative reviews. We select English as the source language and the other three as the target languages. Therefore, we can evaluate our approach on nine tasks in total (i.e., combinations of the three domains and three target languages). For each task, we use the 2,000 labeled English reviews in the corresponding domain for training and the 2,000 labeled reviews in each target language for evaluation. The translations of the test reviews are already provided in this dataset, so we only need to translate the English reviews into target languages. 

To achieve unlabeled data ($U_S$ and $U_T$), we collect a sample of English, Japanese, French, and German Tweets between September 2016 and March 2018. All collected Tweets are used to train the word embeddings. As emojis are widely used on Twitter~\cite{PavalanathanE15a}, we are able to extract emoji-labeled Tweets, which are used to learn emoji-powered sentence representations. For each language, we extract Tweets containing the top 64 emojis used in this language. 
As many Tweets contain multiple emojis, for each Tweet, we create separate examples for each unique emoji used in it to make the emoji prediction a single-label classification task instead of more complicated multi-label classification. 

We then conduct the following preprocessing procedures for the documents. We remove all Retweets, and Tweets that contain URLs, to ensure that words appear in their original contexts and that the meaning of the Tweets do not depend on external content. Then we tokenize all the texts (including reviews and Tweets) into words, convert them into lowercase, and shorten the words with redundant characters into their canonical forms (e.g., ``cooooool'' is converted to ``cool''). As Japanese words are not separated by white spaces, we use a tokenization tool called \textit{MeCab}~\cite{MeCab} to segment Japanese documents. In addition, we use special tokens to replace mentions and numbers. The processed emoji-Tweets provide the $E_S$ and $E_T$ datasets, whose statistics are presented in Table ~\ref{data_size}.

\subsection{Implementation Details}
We learn the initial word embeddings using the skip-gram model with the window-size of 5 on the raw Tweets. The word vectors are then fine-tuned during the sentence representation learning phase. In the representation learning phase, to regularize our model, L2 regularization with parameter $10^{-6}$ is applied for embedding weights. 
Dropout is applied at the rate of 0.5 before the softmax layer. The hidden units of bi-directional LSTM layers are set as 1,024 (512 in each direction). We randomly split the emoji-Tweets into the training, validation, and test sets in the proportion of 7:2:1. Accordingly, we use early stopping ~\cite{CaruanaLG00} to tune hyperparameters based on the validation performance through 50 epochs, with mini-batch size of 250. We used the Adam algorithm ~\cite{KingmaB14} for optimization, with the two momentum parameters set to 0.9 and 0.999, respectively. The initial learning rate was set to $10^{-3}$. In the phase of training the sentiment classifier, for exhaustive parameter tuning, we randomly select 90\% of the labeled data as the training set and the remaining 10\% as the validation set. The whole framework is implemented with TensorFlow~\cite{abadi2016tensorflow}.

\subsection{Baselines and Accuracy Comparison}
To evaluate the performance of \approach, we employ three representative baseline methods for comparison: 

\textbf{MT-BOW} uses the bag-of-words features to learn a linear classifier on the labeled English data~\cite{PrettenhoferS10}. It uses \textit{Google Translate} to translate the test data into English and applies the pre-trained classifier to predict the sentiment polarity of the translated documents.

\textbf{CL-RL} is the word-aligned representation learning method proposed by Xiao and Guo~\cite{XiaoG13}. It constructs a unified word representation that consists of both language-specific components and shared components, for the source and the target languages. To establish connections between the two languages, it leverages \textit{Google Translate} to create a set of critical parallel word pairs, and then it forces each parallel word pair to share the same word representation. The document representation is computed by taking the average over all words in the document. Given the representation as features, it trains a linear SVM model using the labeled English data.

\textbf{BiDRL} is the document-aligned representation learning method proposed by Zhou \textit{et al.}~\cite{ZhouWX16}. It uses \textit{Google Translate} to create labeled parallel documents and forces the pseudo parallel documents to share the same embedding space. It also enforces constraints to make the document vectors associated with different sentiments fall into different positions in the embedding space. Furthermore, it forces documents with large textual differences but the same sentiment to have similar representations. After this representation learning process, it concatenates the vectors of one document in both languages and trains a logistic regression sentiment classifier.

 As the benchmark datasets have quite balanced positive and negative reviews, we follow the aforementioned studies to use accuracy as an evaluation metric. All the baseline methods have been evaluated with exactly the same training and test data sets used in previous studies~\cite{ZhouWX16}, so we make direct comparisons with their reported results. Unfortunately, we cannot obtain the individual predictions of these methods, so we are not able to report the statistical significance (such as McNemar's test~\cite{dietterich1998approximate}) of the difference between these baselines and \approach. To alleviate this problem and get robust results, we run \approach 10 times with different random initiations and summarize its average accuracy and standard deviation in Table~\ref{overall_result}, as well as the reported performance of the baselines.

\begin{table}[!tp]
\centering
\footnotesize
\caption{The accuracy of \approach (standard deviations in parentheses) and baseline methods on the nine benchmark tasks.}
\label{overall_result}
\begin{tabular}{cc|cccc}
\hline
Language & Domain & MT-BOW &  CL-RL & BiDRL  & \approach\\
\hline
\multirow{3}{*}{Japanese} & Book & 0.702 &  0.711 & 0.732 &\textbf{0.783 (0.003)}\\
 & DVD & 0.713 & 0.731 & 0.768 &\textbf{0.791 (0.004)}\\
 & Music & 0.720 & 0.744 & 0.788 &\textbf{0.808 (0.005)}\\
\hline
\multirow{3}{*}{French} & Book & 0.808 & 0.783 & 0.844 &\textbf{0.860 (0.002)}\\
 & DVD & 0.788 & 0.748 & 0.836 &\textbf{0.857 (0.002)} \\
 & Music & 0.758 & 0.787 & 0.825 &\textbf{0.860 (0.002)}\\
\hline
\multirow{3}{*}{German} & Book & 0.797 & 0.799 & 0.841 &\textbf{0.864 (0.001)}\\
 & DVD & 0.779 & 0.771 & 0.841 &\textbf{0.861 (0.001)}\\
 & Music & 0.772 & 0.773 & 0.847 &\textbf{0.878 (0.002)}\\
\hline
\end{tabular}
\end{table}


As illustrated in Table~\ref{overall_result},  \approach outperforms all three baseline methods on all nine tasks. Looking more closely, the performance of all methods in Japanese sentiment classification is worse than in French and German tasks. According to the language systems defined by ISO 639~\cite{iso}, English, French, and German belong to the same language family (i.e., Indo-European), while Japanese belongs to the Japonic family. In other words, French and German are more in common with English, and it is expected to be easier to translate English texts into French and German and transfer the sentiment knowledge from English to them. Therefore, in fact, Japanese tasks are most difficult and none of the previous methods have been able to achieve an accuracy above 0.8. It is encouraging to find that \approach achieves an accuracy of 0.808 on the Japanese music task and an accuracy close to 0.8 (0.791) on the Japanese DVD task. The 0.783 accuracy on the book task is also non-negligible as it improves on the best existing model by almost 7 percent. 
In addition, although the French and German tasks are a little easier than the Japanese ones, none of the existing approaches can achieve an accuracy over 0.85 on any of the six tasks. However, our approach can achieve a mean accuracy higher than 0.85 on all of the six tasks.

Next, we compare the results more thoroughly and further demonstrate the advantages of our approach. As is shown, the representation learning approaches (CL-RL, BiDRL, and \approach) all outperform the shallow method MT-BOW on most tasks. This is reasonable as representation learning approaches embed words into high-dimensional vectors in a continuous semantic space and thus overcome the feature sparsity issue of traditional bag-of-words approaches. Furthermore, we observe that the document-level representation approaches (BiDRL and \approach) outperform the word-level CL-RL. This indicates that incorporating document-level information into representations is more effective than focusing on individual words. Finally, \approach outperforms the BiDRL on all tasks. In order to narrow the linguistic gap, BiDRL leverages only pseudo parallel texts to learn the common sentiment patterns between languages. Besides the pseudo parallel texts, \approach also learns from the emoji usage in both languages. On the one hand, as a ubiquitous emotional signal, emojis are adopted across languages to express common sentiment patterns, which can complement the pseudo parallel corpus. On the other hand, the language-specific patterns of emoji usage help incorporate the language-specific knowledge of sentiments into the representation learning, which can benefit the downstream sentiment classification in the target language. As a next step, we explore the role of emojis in the learning process with a more comprehensive investigation.
\begin{figure*}[!tp]
    \centering
    \subfigure[Clusters of selected words with Word2Vec representations.]{
        \label{fig:cluster_before}
        \includegraphics[width=0.48\textwidth]{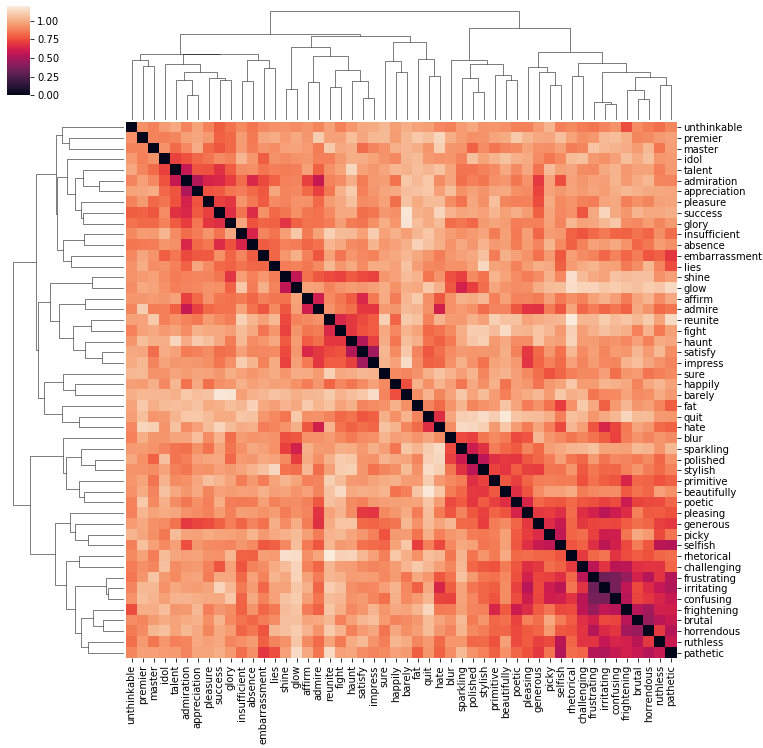}
    }
    \subfigure[Clusters of selected words with emoji-powered representations.]{
        \label{fig:cluster_after}
        \includegraphics[width=0.48\textwidth]{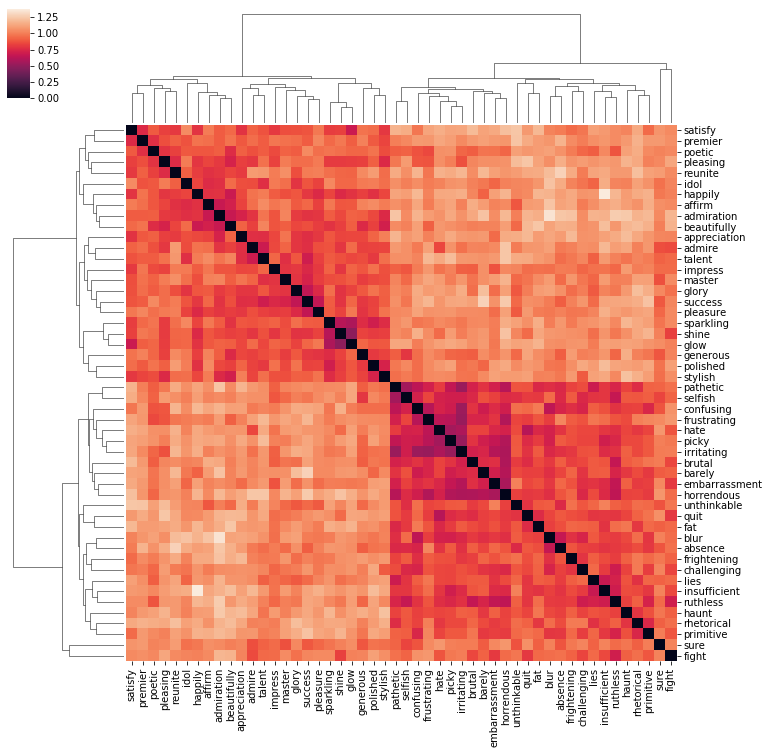}
    }
    \caption{Comparison of word representations \textit{with} and \textit{without} emoji prediction.}\label{cluster_map}
\end{figure*}

\subsection{The Power of Emojis}\label{emoji_effect}
To further evaluate the contribution of emojis in \approach, we conduct subsequent experiments to investigate the effects of emojis from three perspectives, i.e., overall performance, effectiveness of representation learning, and text comprehension. 

\subsubsection{Overall Performance}
To understand how emojis affect cross-lingual sentiment classification in general, a straightforward idea is to remove the emoji-prediction phase and compare simplified versions of \approach:

\textbf{N-\approach} removes the emoji-prediction phase of both languages and directly uses two attention layers to realize the transformation from word vectors to the final document representation. There is no emoji data used in this model.

\textbf{T-\approach} removes the emoji-based representation learning on the English side. It uses the emoji-powered representations for the target language and translates labeled English documents into the target language to train a sentiment classifier for the target language. This model only leverages emoji usage in the target language.

\textbf{S-\approach} removes the emoji-based representation learning in the target language. It uses the emoji-powered representations of English and trains a sentiment classifier based on labeled English documents. Documents in the target language are first translated into English and then classified. This model only leverages emoji usage in the source language (i.e., English).

\begin{table}[!tp]
\begin{threeparttable}
\centering
\small
\caption{Performance of \approach and its simplified versions.}
\label{distant}
\begin{tabular}{cc|cccc}
\hline
Language & Domain & N-\approach & T-\approach & S-\approach & \approach\\
\hline
\multirow{3}{*}{Japanese} & Book & 0.527* & 0.742* & 0.753* & 0.783\\
 & DVD & 0.507* & 0.756* & 0.766* & 0.791\\
 & Music & 0.513* & 0.792* & 0.778* & 0.808\\
\hline
\multirow{3}{*}{French} & Book & 0.505* & 0.821* & 0.850* & 0.860\\
 & DVD & 0.507* & 0.816* & 0.843* & 0.857\\
 & Music & 0.503* & 0.811* & 0.848* & 0.860\\
\hline
\multirow{3}{*}{German} & Book & 0.513* & 0.804* & 0.848* & 0.864\\
 & DVD & 0.521* & 0.790* & 0.849* & 0.861\\
 & Music  & 0.513* & 0.818* & 0.863* & 0.878\\
\hline
\end{tabular}
\begin{tablenotes}
\footnotesize
\item * indicates the difference between \approach and its simplified versions is statistically significant ($p < 0.05$) by McNemar's test.
\end{tablenotes}
\end{threeparttable}
\end{table}

\begin{figure*}
\centering
\subfigure[Word and sentence attention distribution generated by N-\approach.]
{
\begin{minipage}{0.9\linewidth}
\includegraphics[width=15cm]{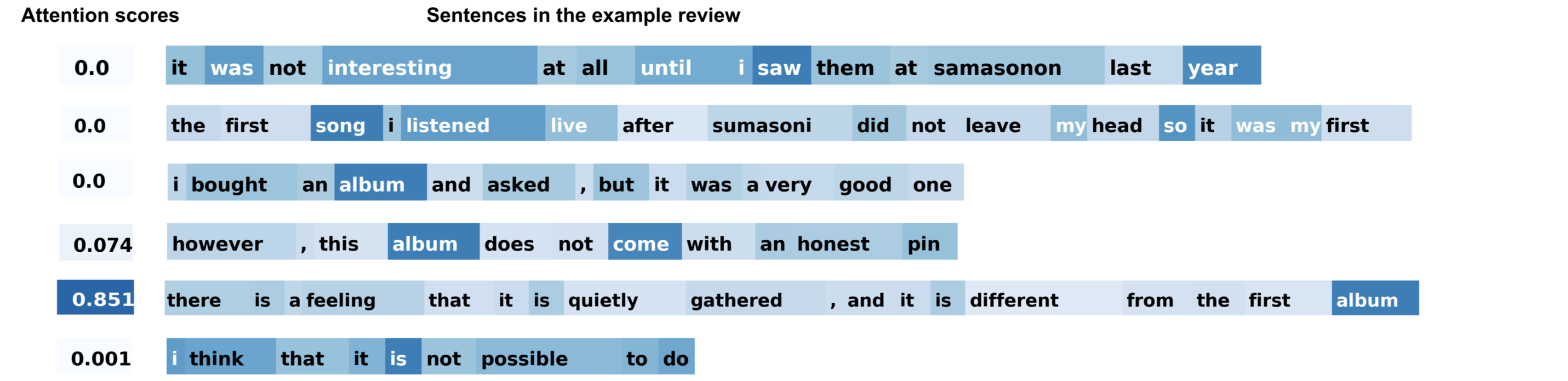}
\label{attention_a}
\end{minipage}
}

\subfigure[Word and sentence attention distribution generated by \approach.]
{
\begin{minipage}{0.9\linewidth}
\includegraphics[width=15cm]{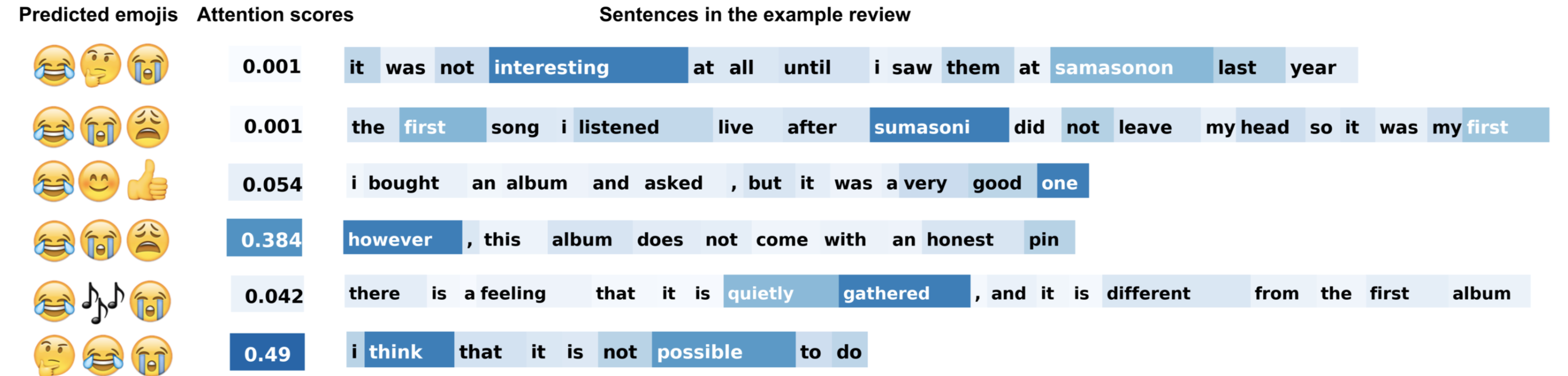}
\label{attention_b}
\end{minipage}
}
\caption{Case study: Effect of emojis on text comprehension.}
\label{fig:attention}
\end{figure*}

Test accuracy of these models is illustrated in Table~\ref{distant}. We find that \approach outperforms N-\approach on all nine tasks. N-\approach is only a little better than uniform guess (50\%) since it learns the common patterns between languages only from pseudo parallel texts and does not incorporate sentiment information effectively. An alternative conjecture is that 2,000 reviews are insufficient to train such a complex model, which may have led to the problem of over-fitting. 

To test between the two hypotheses, we mix up the labeled reviews in English and in the target language and randomly select 2,000 examples from the mixed set for training and use the remaining samples as a new test set. All other settings of the experiment are kept the same except for the new train/test split. Trained and tested in this way, the accuracy of N-\approach becomes acceptable, with an average accuracy of 0.777 on all tasks. This indicates that over-fitting might not have been the major reason, while language discrepancy might be. Indeed, N-\approach can still work well if we effectively incorporate cross-language sentiment information into the training process. More specifically, the original N-\approach is dominated by English sentiment information learned from pseudo parallel texts and fails to generalize to the target language correctly. When we input the sentiment information (labeled documents) of both English and the target language into the model, performance improves. Unfortunately, in a cross-lingual sentiment classification setting, we can not acquire enough labels in the target language. Emojis help the model capture generalizable sentiment knowledge, even if there is no labeled example for training in the target language. 

In addition, \approach also consistently achieves better accuracy compared to T-\approach and S-\approach on all tasks (McNemar's test~\cite{dietterich1998approximate} is performed and the differences are all statistically significant at the 5\% level). 
The superiority of \approach shows that only extracting sentiment information from one language is not enough for the cross-lingual sentiment task and that incorporating language-specific knowledge for both languages is critical to the model's performance. Indeed, S-\approach fails to capture sentiment patterns in the target language; and T-\approach falls short in extracting transferable sentiment patterns from English (indicating that emojis are still beneficial even if there are sentiment labels in a language). 

\begin{figure*}[!tp]
    \centering
    \subfigure[Unlabeled data: Japanese tasks]{
        \label{fig:udataj}
        \includegraphics[width=0.27\textwidth]{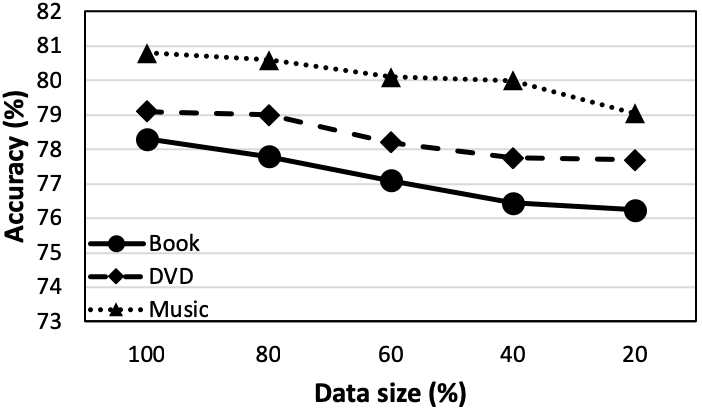}
    }
    \subfigure[Unlabeled data: French tasks]{
        \label{fig:udataf}
        \includegraphics[width=0.27\textwidth]{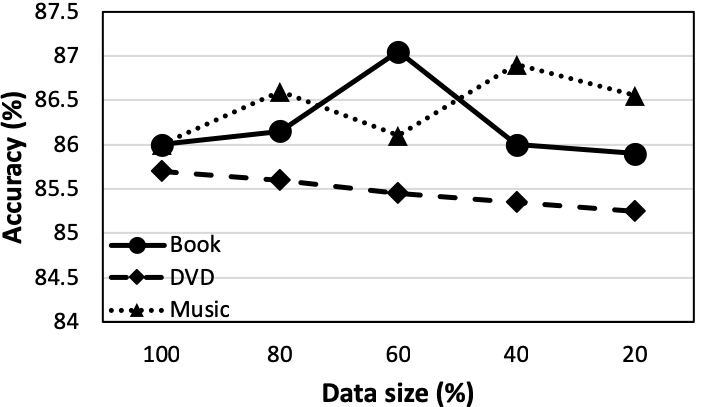}
    }
  	\subfigure[Unlabeled data: German tasks]{
  		\label{fig:udatag}
        \includegraphics[width=0.27\textwidth]{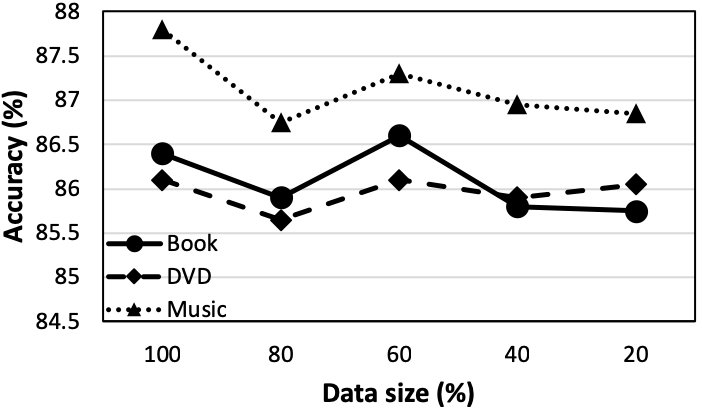}
  	}
    \subfigure[Labeled English data: Japanese tasks]{
        \label{fig:dataj}
        \includegraphics[width=0.27\textwidth]{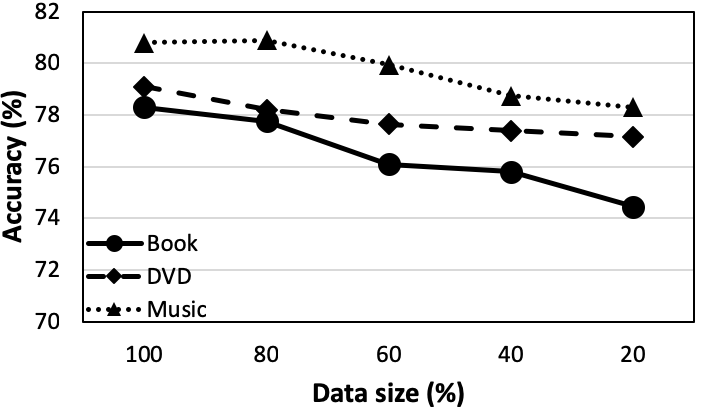}
    }
    \subfigure[Labeled English data: French tasks]{
        \label{fig:dataf}
        \includegraphics[width=0.27\textwidth]{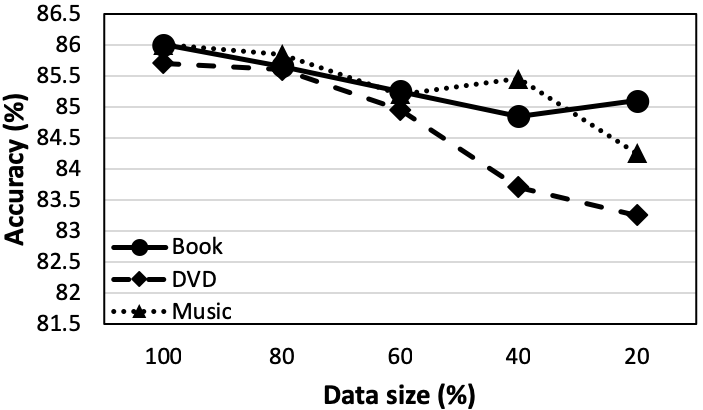}
    }
  	\subfigure[Labeled English data: German tasks]{
  		\label{fig:datag}
        \includegraphics[width=0.27\textwidth]{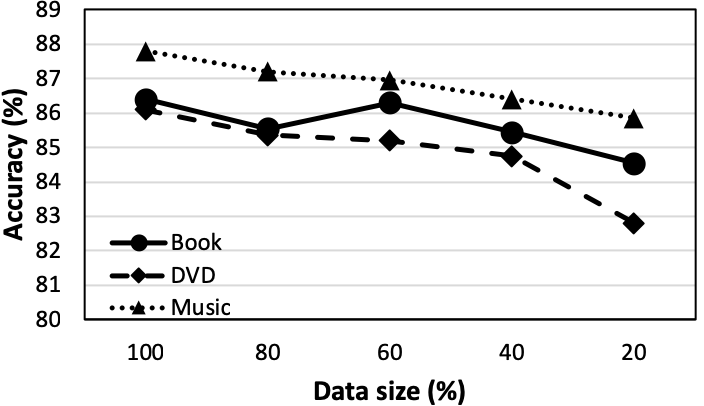}
  	}
    \caption{Accuracy of \approach when size of unlabeled and labeled data changes.}\label{datasize}
\end{figure*}

\subsubsection{Effectiveness of Representation Learning}
To better understand the sentiment information learned through the emoji usage, we then conduct an empirical experiment at the word representation level. Recall that after the word embedding phase, each individual word can be represented by a unique vector and that these word vectors are then fine-tuned in the emoji-prediction phase. Next, we would like to evaluate whether sentiment information is better captured by the new word representations under the effects of emojis. We sample 50 English words with distinct sentiments from the MPQA subjectivity lexicon~\cite{mpqa} based on their frequencies in our corpus. These words are manually labeled in terms of positive or negative polarity from MPQA, and we regard these labels as the ground-truth for further evaluation. 

We expect that an informative representation can embed words with same sentiment polarity closely in the vector space. To measure and illustrate the similarity, we calculate the similarity score between every two words using the cosine of the corresponding embedding vectors.
Based on the cosine similarity, we perform a hierarchical clustering~\cite{JosephGJ01} and visualize the clustering results in Figure~\ref{cluster_map}. The color scale of each cell indicates the similarity between the two words. The darker the cell, the more similar the representations of the two words. 

In Figure~\ref{fig:cluster_before}, we use naive embeddings learned by Word2Vec (no emoji), and words with different sentiments cannot be clearly separated. Many words with different sentiments are embedded closely, for example, ``generous'' and ``picky'' in the bottom right section. This indicates that shallow word embeddings do not effectively capture the sentiment information. 

In contrast, in Figure~\ref{fig:cluster_after}, we can easily observe two clusters after the fine-tuned emoji-prediction model. The top left corner cluster contains the positive words, while the bottom right corner contains the negative words. Only one positive word, ``sure,'' is incorrectly clustered with negative words. By checking the contexts of this word in our corpus, we find it is usually co-used with both positive and negative words, making its polarity ambiguous.
The correct clustering of nearly all the words indicates that emoji usage is an effective channel to capture sentiment knowledge, which is desirable for downstream sentiment classification.

\subsubsection{Text Comprehension}
We then explore how the emoji-powered representations benefit text comprehension. We select a representative case that is incorrectly classified by N-\approach but correctly classified by \approach. This case is selected from the Japanese test samples and we use the segment of its translated English version for illustration in Figure~\ref{fig:attention}. Although the whole document expresses dissatisfaction with an album, it is not that easy to identify this intent directly from each single sentence due to the translation quality and the document's complex compositions. For example, if we consider only the third sentence without context, the author seems to express a positive attitude. However, in fact, the author expresses an obviously negative attitude in the fourth and sixth sentences.

In Figure~\ref{fig:attention}, we present the attention distribution of words and sentences generated by N-\approach and \approach, which indicates how the two models comprehend this document, or the rationale behind their classification decisions. We use the color scale of the background to indicate the attention scores of words in each sentence. The darker the word, the more it is attended to. For each sentence, we list its attention score in this document. In Figure~\ref{attention_b}, we also list the top 3 emojis \approach predicts for each sentence, which may indicate its sentiment polarity predicted by \approach.  

Let us first look at Figure~\ref{attention_a}, which demonstrates how N-\approach processes the sentiment information. On the word level, N-\approach tends to focus more on neutral words like ``song'' or ``album'' instead of sentimental words. On the sentence level, an extremely high attention is placed on the fifth sentence. However, the fifth sentence describes how the album is different from the first one and it does not express the obviously negative sentiment.

In contrast, after incorporating of emojis, \approach is able to work with a proper logic (see Figure~\ref{attention_b}). 
\approach places its attention to the emotional adjectives, such as ``interesting'' and ``not possible,'' and contrast conjunctions such as ``however.'' Thus, it manages to identify the sentiment of each sentence as expected, which can be further explained by the predicted emojis on the left. Besides the most popular \emoji{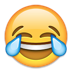} in our corpus, \emoji{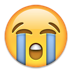} and \emoji{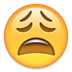} predicted for the fourth and sixth sentence indicate the negative sentiment of the author, while \emoji{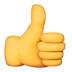} and \emoji{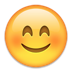} in the third sentence indicate positive sentiment. Then on the sentence level, \approach places less attention to the positive third sentence, while centering upon the fourth and the sixth sentences. Through this comparison, we can see that emojis bring additional knowledge to the text comprehension and make the attention mechanism more effective. 
\section{Discussion}\label{discussion}
So far, we have presented the performance of \approach on benchmark datasets and demonstrated the power of emojis in our learning process. There are some issues that could potentially affect its effectiveness and efficiency, which call for further discussion. 

\subsection{Sensitivity on Data Volume}
As we learn text representations from large amount of Tweets, we want to investigate whether \approach works well with a smaller volume of data. First, we investigate the size of unlabeled data. The English representation model, once learned, can be reused by any other English-target language pair. We only need to scale down the Tweets and emoji-Tweets in the target language and observe the changes in performance on benchmarks. In details, we use 80\%, 60\%, 40\%, and 20\% of the collected Tweets to re-train the target-language representation model and keep the final supervised training fixed. We summarize the results in Figures~\ref{fig:udataj},~\ref{fig:udataf}, and~\ref{fig:udatag}. For the Japanese tasks, when we scale down the unlabeled data, the performance gets slightly worse. Comparing the results using 20\% and 100\% of the Tweets, the accuracy differences in three domains are 0.021, 0.014, and 0.018, respectively. 
For French and German, the performance fluctuates less than 0.01. Most importantly, \approach can outperform the existing methods on all nine tasks even with the 20\% unlabeled data. This indicates that even though a target language is not as actively used on Twitter, our approach still works.

Furthermore, although there are more labeled examples in English than other languages, in general, labels are still scarce. Hence, if a model can rely on even fewer labeled English documents, it is very desirable. To test this, we scale down the labeled data by 80\%, 60\%, 40\%, and 20\%. As shown in Figures~\ref{fig:dataj},~\ref{fig:dataf}, and~\ref{fig:datag}, the performance of \approach slightly declines with the decrease of labels, 
but even with 20\% labels (i.e., 400 labeled English samples), \approach outperforms the existing methods using all 2,000 labeled samples on almost all tasks. This shows that with the help of large-scale emoji-Tweets, the model is less dependent on sentiment labels.

\subsection{Generalizability}
Most previous cross-lingual sentiment studies~\cite{PrettenhoferS10,XiaoG13,ZhouWX16} used the Amazon review dataset for evaluation. To compare with them, we also adopt this dataset in the main experiment of this paper.
Sentiment classification in other domains such as social media is also important. Can \approach still work well in a new domain? To evaluate the generalizability of our approach, we apply \approach to a representative type of social media data -- Tweets. As Tweets are short and informal, sentiment classification for them is considered to be a big challenge~\cite{GiachanouC16}. 

As cross-lingual studies on Tweets are very limited, we take only one recent cross-lingual method (MT-CNN) proposed by Deriu \textit{et al.}~\cite{DeriuLLSMCHJ17} for comparison. It also relies on large-scale unlabeled Tweets and a translation tool. It first trains a sentiment classifier for English and then applies it to the translations of text documents in the target language. The training process for English Tweets contains three phases. First, it uses raw Tweets to create word embeddings just like our method. Second, it leverages ``:)'' and ``:('' as weak labels and applies a multi-layer CNN model to adapt the word embeddings. Finally, it trains the model on labeled English Tweets. This work and our work both have coverage of French and German Tweets, so we use the two as the target languages for comparison. 

As the sentiment-labeled Tweets used by \cite{DeriuLLSMCHJ17} are released in forms of Twitter IDs and some of them are no longer available now, we cannot directly compare our model to the reported results in ~\cite{DeriuLLSMCHJ17}. For fair comparison, we reproduce their method on the labeled Tweets that can still be collected. Based on the pre-trained representation models of MT-CNN~\cite{mlsa} and \approach, we use the same labeled English Tweets to train and validate the two classifiers and then test them on the same data (i.e., labeled French and German Tweets that can be collected). We list the sizes of the labeled English, French, and German Tweets we use in Table~\ref{distribution}. From the distribution, a naive baseline using uniform guess would achieve an accuracy of 0.451 for French and 0.628 for German. 

\begin{table}[!tp]
\centering
\small
\caption{The sizes of labeled Tweets collected.}
\label{distribution}
\begin{tabular}{ll|rrr}
\hline
Dataset & Language & Positive & Neutral & Negative\\
\hline
Training & English~\cite{entrain} & 5,101 & 3,742 & 1,643 \\
\hline
Validation & English~\cite{entest} & 1,038 & 987 & 365 \\
\hline
\multirow{2}{*}{Test} & French~\cite{frenchsenti} & 987 & 1,389 & 718\\
& German~\cite{germansenti} & 1,057 & 4,441 & 1,573 \\
\hline
\end{tabular}
\end{table}

\begin{table}[!tp]
\begin{threeparttable}
\centering
\small
\caption{Classification accuracy on French and German Tweets.}
\label{Tweet_result}
\begin{tabular}{l|ccc}
\hline
Language & \approach & MT-CNN & Uniform Guess\\
\hline
French & 0.696  & 0.535* & 0.451* \\
\hline
German & 0.809 & 0.654* & 0.628* \\
\hline
\end{tabular}
\begin{tablenotes}
\footnotesize
\item * indicates the difference between \approach and the baseline methods is statistically significant ($p < 0.05$) by McNemar's test.
\end{tablenotes}
\end{threeparttable}
\end{table}

Results are summarized in Table~\ref{Tweet_result}. The two approaches both outperform uniform guess, and \approach outperforms the MT-CNN by 0.161 on French task and 0.155 on German task. Although we use the same training, validation, and test set for both approaches, we are still concerned about whether the pre-trained representation models have introduced unfairness. Specifically, if we have used more unlabeled Tweets for representation learning than MT-CNN, our outstanding performance may simply attribute to the size of data. To answer this question, we refer to ~\cite{DeriuLLSMCHJ17} about their data size. We find that they uses 300M raw Tweets and 60M Tweets containing ``:)'' and ``:('' for representation learning. In contrast, we only used 81M raw Tweets and 13.7M emoji-Tweets in three languages combined. Considering that emoticons are significantly less used than emojis on Twitter~\cite{PavalanathanE15a}, although they use about 4.4 times more weak-labeled Tweets, these Tweets had to be collected from much more than 4.4 times of raw Tweets than ours. It is clear \approach outperforms MT-CNN and relies less on data size.

\section{Conclusion}\label{conclusion}
As a ubiquitous emotional signal, emojis are widely adopted across languages to express sentiments. We leverage this characteristic of emojis, both using them as surrogate sentiment labels and using emoji prediction an instrument to address the language discrepancy in cross-lingual sentiment classification. We have presented \approach, a novel emoji-powered representation learning framework, to capture both general and language-specific sentiment knowledge in the source and the target languages for cross-lingual sentiment classification. 
The representations learned by \approach capture not only sentiment knowledge that generalizes across languages, but also language-specific patterns. 
We evaluate \approach with comprehensive experiments on various benchmark datasets, which outperforms the state-of-the-art cross-lingual sentiment classification methods even when the size of labeled and unlabeled data decreases. The promising results indicate that emojis may be used as an a general instrument for text mining tasks that suffer from the scarcity of labeled examples, especially in situations where an inequality among different languages presents.  
\section*{Acknowledgment}
This work was in part supported by the National Key R\&D Program of China under the grant number 2018YFB1004800 and the Beijing Municipal Science and Technology Project under the grant number Z171100005117002. Qiaozhu Mei's work was supported by the National Science Foundation under grant numbers 1633370, 1131500, and 1620319. The authors would like to thank the invaluable supports from Mr. Wei Ai at University of Michigan and Ms. Jiawei Liu at Peking University. Zhenpeng Chen and Sheng Shen made equal contributions to this work. The work of Sheng Shen and Ziniu Hu was carried out when they were undergraduate students at Peking University. 
\bibliographystyle{ACM-Reference-Format}
\bibliography{emojisentiment-bibliography}


\begin{thebibliography}{62}


\ifx \showCODEN    \undefined \def \showCODEN     #1{\unskip}     \fi
\ifx \showDOI      \undefined \def \showDOI       #1{#1}\fi
\ifx \showISBNx    \undefined \def \showISBNx     #1{\unskip}     \fi
\ifx \showISBNxiii \undefined \def \showISBNxiii  #1{\unskip}     \fi
\ifx \showISSN     \undefined \def \showISSN      #1{\unskip}     \fi
\ifx \showLCCN     \undefined \def \showLCCN      #1{\unskip}     \fi
\ifx \shownote     \undefined \def \shownote      #1{#1}          \fi
\ifx \showarticletitle \undefined \def \showarticletitle #1{#1}   \fi
\ifx \showURL      \undefined \def \showURL       {\relax}        \fi
\providecommand\bibfield[2]{#2}
\providecommand\bibinfo[2]{#2}
\providecommand\natexlab[1]{#1}
\providecommand\showeprint[2][]{arXiv:#2}

\bibitem[\protect\citeauthoryear{??}{mpq}{2005}]%
        {mpqa}
 \bibinfo{year}{2005}\natexlab{}.
\newblock \bibinfo{title}{MPQA opinion corpus}.
\newblock
  \bibinfo{howpublished}{\url{https://mpqa.cs.pitt.edu/lexicons/subj_lexicon/}}.
    (\bibinfo{year}{2005}).
\newblock
\newblock
\shownote{Retrieved on October 22, 2018.}


\bibitem[\protect\citeauthoryear{??}{MeC}{2006}]%
        {MeCab}
 \bibinfo{year}{2006}\natexlab{}.
\newblock \bibinfo{title}{MeCab: Yet Another Part-of-Speech and Morphological
  Analyzer}.
\newblock \bibinfo{howpublished}{\url{http://taku910.github.io/mecab}}.
  (\bibinfo{year}{2006}).
\newblock
\newblock
\shownote{Retrieved on October 22, 2018.}


\bibitem[\protect\citeauthoryear{??}{Web}{2010}]%
        {Webis}
 \bibinfo{year}{2010}\natexlab{}.
\newblock \bibinfo{title}{Webis-CLS-10}.
\newblock
  \bibinfo{howpublished}{\url{https://www.uni-weimar.de/en/media/chairs/computer-science-department/webis/data/corpus-webis-cls-10/}}.
    (\bibinfo{year}{2010}).
\newblock
\newblock
\shownote{Retrieved on October 22, 2018.}


\bibitem[\protect\citeauthoryear{??}{fre}{2015}]%
        {frenchsenti}
 \bibinfo{year}{2015}\natexlab{}.
\newblock \bibinfo{title}{DEFT 2015: Test Corpus}.
\newblock
  \bibinfo{howpublished}{\url{https://deft.limsi.fr/2015/corpus.fr.php?lang=en}}.
    (\bibinfo{year}{2015}).
\newblock
\newblock
\shownote{Retrieved on April 28, 2018.}


\bibitem[\protect\citeauthoryear{??}{mls}{2017}]%
        {mlsa}
 \bibinfo{year}{2017}\natexlab{}.
\newblock \bibinfo{title}{Deep-mlsa}.
\newblock
  \bibinfo{howpublished}{\url{https://github.com/spinningbytes/deep-mlsa}}.
  (\bibinfo{year}{2017}).
\newblock
\newblock
\shownote{Retrieved on October 22, 2018.}


\bibitem[\protect\citeauthoryear{??}{ger}{2017}]%
        {germansenti}
 \bibinfo{year}{2017}\natexlab{}.
\newblock \bibinfo{title}{SB-10k: German Sentiment Corpus}.
\newblock
  \bibinfo{howpublished}{\url{https://www.spinningbytes.com/resources/germansentiment/}}.
    (\bibinfo{year}{2017}).
\newblock
\newblock
\shownote{Retrieved on April 28, 2018.}


\bibitem[\protect\citeauthoryear{??}{ent}{2017a}]%
        {entest}
 \bibinfo{year}{2017}\natexlab{a}.
\newblock \bibinfo{title}{Twitter-2015test-A}.
\newblock
  \bibinfo{howpublished}{\url{http://alt.qcri.org/semeval2017/task4/index.php?id=download-the-full-training-data-for-semeval-2017-task-4}}.
    (\bibinfo{year}{2017}).
\newblock
\newblock
\shownote{Retrieved on April 28, 2018.}


\bibitem[\protect\citeauthoryear{??}{ent}{2017b}]%
        {entrain}
 \bibinfo{year}{2017}\natexlab{b}.
\newblock \bibinfo{title}{Twitter-2015train-A, Twitter-2016train-A,
  Twitter-2016dev-A, and Twitter-2016devtest-A}.
\newblock
  \bibinfo{howpublished}{\url{http://alt.qcri.org/semeval2017/task4/index.php?id=download-the-full-training-data-for-semeval-2017-task-4}}.
    (\bibinfo{year}{2017}).
\newblock
\newblock
\shownote{Retrieved on April 28, 2018.}


\bibitem[\protect\citeauthoryear{??}{wor}{2018}]%
        {worlduser}
 \bibinfo{year}{2018}\natexlab{}.
\newblock \bibinfo{title}{Internet world users by language}.
\newblock
  \bibinfo{howpublished}{\url{https://www.internetworldstats.com/stats7.htm}}.
   (\bibinfo{year}{2018}).
\newblock
\newblock
\shownote{Retrieved on October 22, 2018.}


\bibitem[\protect\citeauthoryear{??}{iso}{2018}]%
        {iso}
 \bibinfo{year}{2018}\natexlab{}.
\newblock \bibinfo{title}{ISO 639}.
\newblock
  \bibinfo{howpublished}{\url{https://www.iso.org/iso-639-language-codes.html}}.
    (\bibinfo{year}{2018}).
\newblock
\newblock
\shownote{Retrieved on October 22, 2018.}


\bibitem[\protect\citeauthoryear{Abadi, Barham, Chen, Chen, Davis, Dean, Devin,
  Ghemawat, Irving, Isard, et~al\mbox{.}}{Abadi et~al\mbox{.}}{2016}]%
        {abadi2016tensorflow}
\bibfield{author}{\bibinfo{person}{Mart{\'\i}n Abadi}, \bibinfo{person}{Paul
  Barham}, \bibinfo{person}{Jianmin Chen}, \bibinfo{person}{Zhifeng Chen},
  \bibinfo{person}{Andy Davis}, \bibinfo{person}{Jeffrey Dean},
  \bibinfo{person}{Matthieu Devin}, \bibinfo{person}{Sanjay Ghemawat},
  \bibinfo{person}{Geoffrey Irving}, \bibinfo{person}{Michael Isard},
  {et~al\mbox{.}}} \bibinfo{year}{2016}\natexlab{}.
\newblock \showarticletitle{Tensorflow: a system for large-scale machine
  learning}. In \bibinfo{booktitle}{\emph{OSDI}}, Vol.~\bibinfo{volume}{16}.
  \bibinfo{pages}{265--283}.
\newblock


\bibitem[\protect\citeauthoryear{Ai, Lu, Liu, Wang, Huang, and Mei}{Ai
  et~al\mbox{.}}{2017}]%
        {AiLLW0M17}
\bibfield{author}{\bibinfo{person}{Wei Ai}, \bibinfo{person}{Xuan Lu},
  \bibinfo{person}{Xuanzhe Liu}, \bibinfo{person}{Ning Wang},
  \bibinfo{person}{Gang Huang}, {and} \bibinfo{person}{Qiaozhu Mei}.}
  \bibinfo{year}{2017}\natexlab{}.
\newblock \showarticletitle{Untangling emoji popularity through semantic
  embeddings}. In \bibinfo{booktitle}{\emph{Proceedings of the Eleventh
  International Conference on Web and Social Media, {ICWSM} 2017}}.
  \bibinfo{pages}{2--11}.
\newblock


\bibitem[\protect\citeauthoryear{Bahdanau, Cho, and Bengio}{Bahdanau
  et~al\mbox{.}}{2014}]%
        {BahdanauCB14}
\bibfield{author}{\bibinfo{person}{Dzmitry Bahdanau},
  \bibinfo{person}{Kyunghyun Cho}, {and} \bibinfo{person}{Yoshua Bengio}.}
  \bibinfo{year}{2014}\natexlab{}.
\newblock \showarticletitle{Neural machine translation by jointly learning to
  align and translate}. In \bibinfo{booktitle}{\emph{Proceedings of the 3rd
  International Conference on Learning Representations, ICLR 2014}}.
\newblock


\bibitem[\protect\citeauthoryear{Balikas, Moura, and Amini}{Balikas
  et~al\mbox{.}}{2017}]%
        {BalikasMA17}
\bibfield{author}{\bibinfo{person}{Georgios Balikas}, \bibinfo{person}{Simon
  Moura}, {and} \bibinfo{person}{Massih{-}Reza Amini}.}
  \bibinfo{year}{2017}\natexlab{}.
\newblock \showarticletitle{Multitask learning for fine-grained Twitter
  sentiment analysis}. In \bibinfo{booktitle}{\emph{Proceedings of the 40th
  International {ACM} {SIGIR} Conference on Research and Development in
  Information Retrieval, SIGIR 2017}}. \bibinfo{pages}{1005--1008}.
\newblock


\bibitem[\protect\citeauthoryear{Bar{-}Joseph, Gifford, and
  Jaakkola}{Bar{-}Joseph et~al\mbox{.}}{2001}]%
        {JosephGJ01}
\bibfield{author}{\bibinfo{person}{Ziv Bar{-}Joseph}, \bibinfo{person}{David~K.
  Gifford}, {and} \bibinfo{person}{Tommi~S. Jaakkola}.}
  \bibinfo{year}{2001}\natexlab{}.
\newblock \showarticletitle{Fast optimal leaf ordering for hierarchical
  clustering}. In \bibinfo{booktitle}{\emph{Proceedings of the Ninth
  International Conference on Intelligent Systems for Molecular Biology}}.
  \bibinfo{pages}{22--29}.
\newblock


\bibitem[\protect\citeauthoryear{Barbieri, Kruszewski, Ronzano, and
  Saggion}{Barbieri et~al\mbox{.}}{2016}]%
        {BarbieriKRS16}
\bibfield{author}{\bibinfo{person}{Francesco Barbieri},
  \bibinfo{person}{Germ{\'{a}}n Kruszewski}, \bibinfo{person}{Francesco
  Ronzano}, {and} \bibinfo{person}{Horacio Saggion}.}
  \bibinfo{year}{2016}\natexlab{}.
\newblock \showarticletitle{How cosmopolitan are emojis?: Exploring emojis
  usage and meaning over different languages with distributional semantics}. In
  \bibinfo{booktitle}{\emph{Proceedings of the 2016 {ACM} Conference on
  Multimedia Conference, {MM} 2016}}. \bibinfo{pages}{531--535}.
\newblock


\bibitem[\protect\citeauthoryear{Caruana, Lawrence, and Giles}{Caruana
  et~al\mbox{.}}{2000}]%
        {CaruanaLG00}
\bibfield{author}{\bibinfo{person}{Rich Caruana}, \bibinfo{person}{Steve
  Lawrence}, {and} \bibinfo{person}{C.~Lee Giles}.}
  \bibinfo{year}{2000}\natexlab{}.
\newblock \showarticletitle{Overfitting in neural nets: backpropagation,
  conjugate gradient, and early stopping}. In
  \bibinfo{booktitle}{\emph{Proceedings of advances in neural information
  processing systems 13, NIPS 2000}}. \bibinfo{pages}{402--408}.
\newblock


\bibitem[\protect\citeauthoryear{Chandar, Lauly, Larochelle, Khapra, Ravindran,
  Raykar, and Saha}{Chandar et~al\mbox{.}}{2014}]%
        {PLLKRRS14}
\bibfield{author}{\bibinfo{person}{A.~P.~Sarath Chandar},
  \bibinfo{person}{Stanislas Lauly}, \bibinfo{person}{Hugo Larochelle},
  \bibinfo{person}{Mitesh~M. Khapra}, \bibinfo{person}{Balaraman Ravindran},
  \bibinfo{person}{Vikas~C. Raykar}, {and} \bibinfo{person}{Amrita Saha}.}
  \bibinfo{year}{2014}\natexlab{}.
\newblock \showarticletitle{An autoencoder approach to learning bilingual word
  representations}. In \bibinfo{booktitle}{\emph{Advances in Neural Information
  Processing Systems 27, NIPS 2014}}. \bibinfo{pages}{1853--1861}.
\newblock


\bibitem[\protect\citeauthoryear{Chen, Li, and Li}{Chen et~al\mbox{.}}{2017}]%
        {ChenLL17}
\bibfield{author}{\bibinfo{person}{Qiang Chen}, \bibinfo{person}{Chenliang Li},
  {and} \bibinfo{person}{Wenjie Li}.} \bibinfo{year}{2017}\natexlab{}.
\newblock \showarticletitle{Modeling language discrepancy for cross-lingual
  sentiment analysis}. In \bibinfo{booktitle}{\emph{Proceedings of the 2017
  {ACM} on Conference on Information and Knowledge Management, {CIKM} 2017}}.
  \bibinfo{pages}{117--126}.
\newblock


\bibitem[\protect\citeauthoryear{Chen, Lu, Ai, Li, Mei, and Liu}{Chen
  et~al\mbox{.}}{2018}]%
        {zhenpeng18}
\bibfield{author}{\bibinfo{person}{Zhenpeng Chen}, \bibinfo{person}{Xuan Lu},
  \bibinfo{person}{Wei Ai}, \bibinfo{person}{Huoran Li},
  \bibinfo{person}{Qiaozhu Mei}, {and} \bibinfo{person}{Xuanzhe Liu}.}
  \bibinfo{year}{2018}\natexlab{}.
\newblock \showarticletitle{Through a gender lens: learning usage patterns of
  emojis from large-scale Android users}. In
  \bibinfo{booktitle}{\emph{Proceedings of the 2018 World Wide Web Conference,
  WWW 2018}}. \bibinfo{pages}{763--772}.
\newblock


\bibitem[\protect\citeauthoryear{Cramer, de~Juan, and Tetreault}{Cramer
  et~al\mbox{.}}{2016}]%
        {CramerJT16}
\bibfield{author}{\bibinfo{person}{Henriette Cramer}, \bibinfo{person}{Paloma
  de Juan}, {and} \bibinfo{person}{Joel~R. Tetreault}.}
  \bibinfo{year}{2016}\natexlab{}.
\newblock \showarticletitle{Sender-intended functions of emojis in {US}
  messaging}. In \bibinfo{booktitle}{\emph{Proceedings of the 18th
  International Conference on Human-Computer Interaction with Mobile Devices
  and Services, MobileHCI 2016}}. \bibinfo{pages}{504--509}.
\newblock


\bibitem[\protect\citeauthoryear{Davidov, Tsur, and Rappoport}{Davidov
  et~al\mbox{.}}{2010}]%
        {DavidovTR10}
\bibfield{author}{\bibinfo{person}{Dmitry Davidov}, \bibinfo{person}{Oren
  Tsur}, {and} \bibinfo{person}{Ari Rappoport}.}
  \bibinfo{year}{2010}\natexlab{}.
\newblock \showarticletitle{Enhanced sentiment learning using Twitter hashtags
  and smileys}. In \bibinfo{booktitle}{\emph{Proceedings of the 23rd
  International Conference on Computational Linguistics, {COLING} 2010}}.
  \bibinfo{pages}{241--249}.
\newblock


\bibitem[\protect\citeauthoryear{Deriu, Lucchi, Luca, Severyn, M{\"{u}}ller,
  Cieliebak, Hofmann, and Jaggi}{Deriu et~al\mbox{.}}{2017}]%
        {DeriuLLSMCHJ17}
\bibfield{author}{\bibinfo{person}{Jan Deriu}, \bibinfo{person}{Aur{\'{e}}lien
  Lucchi}, \bibinfo{person}{Valeria~De Luca}, \bibinfo{person}{Aliaksei
  Severyn}, \bibinfo{person}{Simon M{\"{u}}ller}, \bibinfo{person}{Mark
  Cieliebak}, \bibinfo{person}{Thomas Hofmann}, {and} \bibinfo{person}{Martin
  Jaggi}.} \bibinfo{year}{2017}\natexlab{}.
\newblock \showarticletitle{Leveraging large amounts of weakly supervised data
  for multi-language sentiment classification}. In
  \bibinfo{booktitle}{\emph{Proceedings of the 26th International Conference on
  World Wide Web, {WWW} 2017}}. \bibinfo{pages}{1045--1052}.
\newblock


\bibitem[\protect\citeauthoryear{Diakopoulos and Shamma}{Diakopoulos and
  Shamma}{2010}]%
        {DiakopoulosS10}
\bibfield{author}{\bibinfo{person}{Nicholas Diakopoulos} {and}
  \bibinfo{person}{David~A. Shamma}.} \bibinfo{year}{2010}\natexlab{}.
\newblock \showarticletitle{Characterizing debate performance via aggregated
  Twitter sentiment}. In \bibinfo{booktitle}{\emph{Proceedings of the 28th
  International Conference on Human Factors in Computing Systems, {CHI} 2010}}.
  \bibinfo{pages}{1195--1198}.
\newblock


\bibitem[\protect\citeauthoryear{Dietterich}{Dietterich}{1998}]%
        {dietterich1998approximate}
\bibfield{author}{\bibinfo{person}{Thomas~G Dietterich}.}
  \bibinfo{year}{1998}\natexlab{}.
\newblock \showarticletitle{Approximate statistical tests for comparing
  supervised classification learning algorithms}.
\newblock \bibinfo{journal}{\emph{Neural computation}} \bibinfo{volume}{10},
  \bibinfo{number}{7} (\bibinfo{year}{1998}), \bibinfo{pages}{1895--1923}.
\newblock


\bibitem[\protect\citeauthoryear{Felbo, Mislove, S{\o}gaard, Rahwan, and
  Lehmann}{Felbo et~al\mbox{.}}{2017}]%
        {FelboMSRL17}
\bibfield{author}{\bibinfo{person}{Bjarke Felbo}, \bibinfo{person}{Alan
  Mislove}, \bibinfo{person}{Anders S{\o}gaard}, \bibinfo{person}{Iyad Rahwan},
  {and} \bibinfo{person}{Sune Lehmann}.} \bibinfo{year}{2017}\natexlab{}.
\newblock \showarticletitle{Using millions of emoji occurrences to learn
  any-domain representations for detecting sentiment, emotion and sarcasm}. In
  \bibinfo{booktitle}{\emph{Proceedings of the 2017 Conference on Empirical
  Methods in Natural Language Processing, {EMNLP} 2017}}.
  \bibinfo{pages}{1615--1625}.
\newblock


\bibitem[\protect\citeauthoryear{Gamon}{Gamon}{2004}]%
        {Gamon04a}
\bibfield{author}{\bibinfo{person}{Michael Gamon}.}
  \bibinfo{year}{2004}\natexlab{}.
\newblock \showarticletitle{Sentiment classification on customer feedback data:
  noisy data, large feature vectors, and the role of linguistic analysis}. In
  \bibinfo{booktitle}{\emph{Proceedings of 20th International Conference on
  Computational Linguistics, {COLING} 2004}}.
\newblock


\bibitem[\protect\citeauthoryear{Giachanou and Crestani}{Giachanou and
  Crestani}{2016}]%
        {GiachanouC16}
\bibfield{author}{\bibinfo{person}{Anastasia Giachanou} {and}
  \bibinfo{person}{Fabio Crestani}.} \bibinfo{year}{2016}\natexlab{}.
\newblock \showarticletitle{Like it or not: a survey of Twitter sentiment
  analysis methods}.
\newblock \bibinfo{journal}{\emph{Comput. Surveys}} \bibinfo{volume}{49},
  \bibinfo{number}{2} (\bibinfo{year}{2016}), \bibinfo{pages}{28:1--28:41}.
\newblock


\bibitem[\protect\citeauthoryear{Harakawa, Takehara, Ogawa, and
  Haseyama}{Harakawa et~al\mbox{.}}{2018}]%
        {HarakawaTOH18}
\bibfield{author}{\bibinfo{person}{Ryosuke Harakawa}, \bibinfo{person}{Daichi
  Takehara}, \bibinfo{person}{Takahiro Ogawa}, {and} \bibinfo{person}{Miki
  Haseyama}.} \bibinfo{year}{2018}\natexlab{}.
\newblock \showarticletitle{Sentiment-aware personalized Tweet recommendation
  through multimodal {FFM}}.
\newblock \bibinfo{journal}{\emph{Multimedia Tools Appl.}}
  \bibinfo{volume}{77}, \bibinfo{number}{14} (\bibinfo{year}{2018}),
  \bibinfo{pages}{18741--18759}.
\newblock


\bibitem[\protect\citeauthoryear{Herdem}{Herdem}{2012}]%
        {Herdem12}
\bibfield{author}{\bibinfo{person}{Kiraz~Candan Herdem}.}
  \bibinfo{year}{2012}\natexlab{}.
\newblock \showarticletitle{Reactions: Twitter based mobile application for
  awareness of friends' emotions}. In \bibinfo{booktitle}{\emph{Proceedings of
  the 2012 ACM Conference on Ubiquitous Computing, UbiComp 2012}}.
  \bibinfo{pages}{796--797}.
\newblock


\bibitem[\protect\citeauthoryear{Hermans and Schrauwen}{Hermans and
  Schrauwen}{2013}]%
        {Hermans2013Training}
\bibfield{author}{\bibinfo{person}{M. Hermans} {and} \bibinfo{person}{B.
  Schrauwen}.} \bibinfo{year}{2013}\natexlab{}.
\newblock \showarticletitle{Training and analysing deep recurrent neural
  networks}.
\newblock \bibinfo{journal}{\emph{Proceedings of advances in Neural Information
  Processing Systems, NIPS 2013}}, \bibinfo{pages}{190--198}.
\newblock


\bibitem[\protect\citeauthoryear{Hochreiter}{Hochreiter}{1998}]%
        {Hochreiter98}
\bibfield{author}{\bibinfo{person}{Sepp Hochreiter}.}
  \bibinfo{year}{1998}\natexlab{}.
\newblock \showarticletitle{The vanishing gradient problem during learning
  recurrent neural nets and problem solutions}.
\newblock \bibinfo{journal}{\emph{International Journal of Uncertainty,
  Fuzziness and Knowledge-Based Systems}} \bibinfo{volume}{6},
  \bibinfo{number}{2} (\bibinfo{year}{1998}), \bibinfo{pages}{107--116}.
\newblock


\bibitem[\protect\citeauthoryear{Hochreiter and Schmidhuber}{Hochreiter and
  Schmidhuber}{1997}]%
        {Hochreiter1997Long}
\bibfield{author}{\bibinfo{person}{Sepp Hochreiter} {and}
  \bibinfo{person}{Jürgen Schmidhuber}.} \bibinfo{year}{1997}\natexlab{}.
\newblock \showarticletitle{Long short-term memory.}
\newblock \bibinfo{journal}{\emph{Neural Computation}} \bibinfo{volume}{9},
  \bibinfo{number}{8} (\bibinfo{year}{1997}), \bibinfo{pages}{1735--1780}.
\newblock


\bibitem[\protect\citeauthoryear{Hu, Guo, Sun, Nguyen, and Luo}{Hu
  et~al\mbox{.}}{2017}]%
        {HuGSNL17}
\bibfield{author}{\bibinfo{person}{Tianran Hu}, \bibinfo{person}{Han Guo},
  \bibinfo{person}{Hao Sun}, \bibinfo{person}{Thuy{-}vy~Thi Nguyen}, {and}
  \bibinfo{person}{Jiebo Luo}.} \bibinfo{year}{2017}\natexlab{}.
\newblock \showarticletitle{Spice up your chat: the intentions and sentiment
  effects of using emojis}. In \bibinfo{booktitle}{\emph{Proceedings of the
  Eleventh International Conference on Web and Social Media, {ICWSM} 2017}}.
  \bibinfo{pages}{102--111}.
\newblock


\bibitem[\protect\citeauthoryear{Kingma and Ba}{Kingma and Ba}{2014}]%
        {KingmaB14}
\bibfield{author}{\bibinfo{person}{Diederik~P. Kingma} {and}
  \bibinfo{person}{Jimmy Ba}.} \bibinfo{year}{2014}\natexlab{}.
\newblock \showarticletitle{Adam: {a} method for stochastic optimization}.
\newblock \bibinfo{journal}{\emph{CoRR}}  \bibinfo{volume}{abs/1412.6980}
  (\bibinfo{year}{2014}).
\newblock


\bibitem[\protect\citeauthoryear{Ling, Xue, Dai, Jiang, Yang, and Yu}{Ling
  et~al\mbox{.}}{2008}]%
        {LingXDJYY08}
\bibfield{author}{\bibinfo{person}{Xiao Ling}, \bibinfo{person}{Gui{-}Rong
  Xue}, \bibinfo{person}{Wenyuan Dai}, \bibinfo{person}{Yun Jiang},
  \bibinfo{person}{Qiang Yang}, {and} \bibinfo{person}{Yong Yu}.}
  \bibinfo{year}{2008}\natexlab{}.
\newblock \showarticletitle{Can Chinese Web pages be classified with English
  data source?}. In \bibinfo{booktitle}{\emph{Proceedings of the 17th
  International Conference on World Wide Web, {WWW} 2008}}.
  \bibinfo{pages}{969--978}.
\newblock


\bibitem[\protect\citeauthoryear{Liu}{Liu}{2012}]%
        {2012Liu}
\bibfield{author}{\bibinfo{person}{Bing Liu}.} \bibinfo{year}{2012}\natexlab{}.
\newblock \bibinfo{booktitle}{\emph{Sentiment analysis and opinion mining}}.
\newblock \bibinfo{publisher}{Morgan {\&} Claypool Publishers}.
\newblock


\bibitem[\protect\citeauthoryear{Liu, Zhang, Zeng, Huang, and Wu}{Liu
  et~al\mbox{.}}{2018}]%
        {LiuZZHW18}
\bibfield{author}{\bibinfo{person}{Qiao Liu}, \bibinfo{person}{Haibin Zhang},
  \bibinfo{person}{Yifu Zeng}, \bibinfo{person}{Ziqi Huang}, {and}
  \bibinfo{person}{Zufeng Wu}.} \bibinfo{year}{2018}\natexlab{}.
\newblock \showarticletitle{Content attention model for aspect based sentiment
  analysis}. In \bibinfo{booktitle}{\emph{Proceedings of the 2018 World Wide
  Web Conference, {WWW} 2018}}. \bibinfo{pages}{1023--1032}.
\newblock


\bibitem[\protect\citeauthoryear{Liu, Huang, An, and Yu}{Liu
  et~al\mbox{.}}{2007}]%
        {LiuHAY07}
\bibfield{author}{\bibinfo{person}{Yang Liu}, \bibinfo{person}{Xiangji Huang},
  \bibinfo{person}{Aijun An}, {and} \bibinfo{person}{Xiaohui Yu}.}
  \bibinfo{year}{2007}\natexlab{}.
\newblock \showarticletitle{{ARSA:} a sentiment-aware model for predicting
  sales performance using blogs}. In \bibinfo{booktitle}{\emph{Proceedings of
  the 30th Annual International {ACM} {SIGIR} Conference on Research and
  Development in Information Retrieval, SIGIR 2007}}.
  \bibinfo{pages}{607--614}.
\newblock


\bibitem[\protect\citeauthoryear{Lu, Ai, Liu, Li, Wang, Huang, and Mei}{Lu
  et~al\mbox{.}}{2016}]%
        {Lu16}
\bibfield{author}{\bibinfo{person}{Xuan Lu}, \bibinfo{person}{Wei Ai},
  \bibinfo{person}{Xuanzhe Liu}, \bibinfo{person}{Qian Li},
  \bibinfo{person}{Ning Wang}, \bibinfo{person}{Gang Huang}, {and}
  \bibinfo{person}{Qiaozhu Mei}.} \bibinfo{year}{2016}\natexlab{}.
\newblock \showarticletitle{Learning from the ubiquitous language: an empirical
  analysis of emoji usage of smartphone users}. In
  \bibinfo{booktitle}{\emph{Proceedings of the 2016 {ACM} International Joint
  Conference on Pervasive and Ubiquitous Computing, UbiComp 2016}}.
  \bibinfo{pages}{770--780}.
\newblock


\bibitem[\protect\citeauthoryear{McGlohon, Glance, and Reiter}{McGlohon
  et~al\mbox{.}}{2010}]%
        {McGlohonGR10}
\bibfield{author}{\bibinfo{person}{Mary McGlohon}, \bibinfo{person}{Natalie~S.
  Glance}, {and} \bibinfo{person}{Zach Reiter}.}
  \bibinfo{year}{2010}\natexlab{}.
\newblock \showarticletitle{Star quality: aggregating reviews to rank products
  and merchants}. In \bibinfo{booktitle}{\emph{Proceedings of the Fourth
  International Conference on Weblogs and Social Media, {ICWSM} 2010}}.
\newblock


\bibitem[\protect\citeauthoryear{Mikolov, Chen, Corrado, and Dean}{Mikolov
  et~al\mbox{.}}{2013}]%
        {Mikolov2013Efficient}
\bibfield{author}{\bibinfo{person}{Tomas Mikolov}, \bibinfo{person}{Kai Chen},
  \bibinfo{person}{Greg Corrado}, {and} \bibinfo{person}{Jeffrey Dean}.}
  \bibinfo{year}{2013}\natexlab{}.
\newblock \showarticletitle{Efficient estimation of word representations in
  vector space}.
\newblock \bibinfo{journal}{\emph{Computer Science}} (\bibinfo{year}{2013}).
\newblock


\bibitem[\protect\citeauthoryear{Miller, Thebault{-}Spieker, Chang, Johnson,
  Terveen, and Hecht}{Miller et~al\mbox{.}}{2016}]%
        {MillerTCJTH16}
\bibfield{author}{\bibinfo{person}{Hannah~Jean Miller}, \bibinfo{person}{Jacob
  Thebault{-}Spieker}, \bibinfo{person}{Shuo Chang}, \bibinfo{person}{Isaac~L.
  Johnson}, \bibinfo{person}{Loren~G. Terveen}, {and} \bibinfo{person}{Brent~J.
  Hecht}.} \bibinfo{year}{2016}\natexlab{}.
\newblock \showarticletitle{"Blissfully happy" or "ready to fight": varying
  interpretations of emoji}. In \bibinfo{booktitle}{\emph{Proceedings of the
  Tenth International Conference on Web and Social Media, ICWSM 2016}}.
  \bibinfo{pages}{259--268}.
\newblock


\bibitem[\protect\citeauthoryear{Mohammad, Salameh, and Kiritchenko}{Mohammad
  et~al\mbox{.}}{2016}]%
        {MohammadSK16}
\bibfield{author}{\bibinfo{person}{Saif~M. Mohammad}, \bibinfo{person}{Mohammad
  Salameh}, {and} \bibinfo{person}{Svetlana Kiritchenko}.}
  \bibinfo{year}{2016}\natexlab{}.
\newblock \showarticletitle{How translation alters sentiment}.
\newblock \bibinfo{journal}{\emph{J. Artif. Intell. Res.}}
  \bibinfo{volume}{55} (\bibinfo{year}{2016}), \bibinfo{pages}{95--130}.
\newblock


\bibitem[\protect\citeauthoryear{O'Connor, Balasubramanyan, Routledge, and
  Smith}{O'Connor et~al\mbox{.}}{2010}]%
        {OConnorBRS10}
\bibfield{author}{\bibinfo{person}{Brendan O'Connor}, \bibinfo{person}{Ramnath
  Balasubramanyan}, \bibinfo{person}{Bryan~R. Routledge}, {and}
  \bibinfo{person}{Noah~A. Smith}.} \bibinfo{year}{2010}\natexlab{}.
\newblock \showarticletitle{From Tweets to polls: linking text sentiment to
  public opinion time series}. In \bibinfo{booktitle}{\emph{Proceedings of the
  Fourth International Conference on Weblogs and Social Media, {ICWSM} 2010}}.
\newblock


\bibitem[\protect\citeauthoryear{Pang, Lee, and Vaithyanathan}{Pang
  et~al\mbox{.}}{2002}]%
        {pang2002thumbs}
\bibfield{author}{\bibinfo{person}{Bo Pang}, \bibinfo{person}{Lillian Lee},
  {and} \bibinfo{person}{Shivakumar Vaithyanathan}.}
  \bibinfo{year}{2002}\natexlab{}.
\newblock \showarticletitle{Thumbs up?: Sentiment classification using machine
  learning techniques}. In \bibinfo{booktitle}{\emph{Proceedings of the 2002
  Conference on Empirical Methods in Natural Language Processing, {EMNLP}
  2002}}. \bibinfo{pages}{79--86}.
\newblock


\bibitem[\protect\citeauthoryear{Pascanu, Mikolov, and Bengio}{Pascanu
  et~al\mbox{.}}{2013}]%
        {PascanuMB13}
\bibfield{author}{\bibinfo{person}{Razvan Pascanu}, \bibinfo{person}{Tomas
  Mikolov}, {and} \bibinfo{person}{Yoshua Bengio}.}
  \bibinfo{year}{2013}\natexlab{}.
\newblock \showarticletitle{On the difficulty of training recurrent neural
  networks}. In \bibinfo{booktitle}{\emph{Proceedings of the 30th International
  Conference on Machine Learning, {ICML} 2013}}. \bibinfo{pages}{1310--1318}.
\newblock


\bibitem[\protect\citeauthoryear{Pavalanathan and Eisenstein}{Pavalanathan and
  Eisenstein}{2015}]%
        {PavalanathanE15a}
\bibfield{author}{\bibinfo{person}{Umashanthi Pavalanathan} {and}
  \bibinfo{person}{Jacob Eisenstein}.} \bibinfo{year}{2015}\natexlab{}.
\newblock \showarticletitle{Emoticons vs. emojis on Twitter: a causal inference
  approach}.
\newblock \bibinfo{journal}{\emph{CoRR}}  \bibinfo{volume}{abs/1510.08480}
  (\bibinfo{year}{2015}).
\newblock


\bibitem[\protect\citeauthoryear{Pennebaker, Francis, and Booth}{Pennebaker
  et~al\mbox{.}}{1999}]%
        {Pennebaker1999Linguistic}
\bibfield{author}{\bibinfo{person}{J.~W. Pennebaker}, \bibinfo{person}{L.~E.
  Francis}, {and} \bibinfo{person}{R.~J. Booth}.}
  \bibinfo{year}{1999}\natexlab{}.
\newblock \showarticletitle{Linguistic inquiry and word count: LIWC}.
\newblock \bibinfo{journal}{\emph{Lawrence Erlbaum Associates Mahwah Nj}}
  (\bibinfo{year}{1999}).
\newblock


\bibitem[\protect\citeauthoryear{Prettenhofer and Stein}{Prettenhofer and
  Stein}{2010}]%
        {PrettenhoferS10}
\bibfield{author}{\bibinfo{person}{Peter Prettenhofer} {and}
  \bibinfo{person}{Benno Stein}.} \bibinfo{year}{2010}\natexlab{}.
\newblock \showarticletitle{Cross-language text classification using structural
  correspondence learning}. In \bibinfo{booktitle}{\emph{Proceedings of the
  48th Annual Meeting of the Association for Computational Linguistics, ACL
  2010}}. \bibinfo{pages}{1118--1127}.
\newblock


\bibitem[\protect\citeauthoryear{Ptaszynski, Rzepka, Araki, and
  Momouchi}{Ptaszynski et~al\mbox{.}}{2012}]%
        {PtaszynskiRAM12}
\bibfield{author}{\bibinfo{person}{Michal Ptaszynski}, \bibinfo{person}{Rafal
  Rzepka}, \bibinfo{person}{Kenji Araki}, {and} \bibinfo{person}{Yoshio
  Momouchi}.} \bibinfo{year}{2012}\natexlab{}.
\newblock \showarticletitle{Automatically annotating {a} five-billion-word
  corpus of Japanese blogs for affect and sentiment analysis}. In
  \bibinfo{booktitle}{\emph{Proceedings of the 3rd Workshop in Computational
  Approaches to Subjectivity and Sentiment Analysis, WASSA@ACL 2012}}.
  \bibinfo{pages}{89--98}.
\newblock


\bibitem[\protect\citeauthoryear{Qiu, He, Zhang, Shi, Bu, and Chen}{Qiu
  et~al\mbox{.}}{2010}]%
        {QiuHZSBC10}
\bibfield{author}{\bibinfo{person}{Guang Qiu}, \bibinfo{person}{Xiaofei He},
  \bibinfo{person}{Feng Zhang}, \bibinfo{person}{Yuan Shi},
  \bibinfo{person}{Jiajun Bu}, {and} \bibinfo{person}{Chun Chen}.}
  \bibinfo{year}{2010}\natexlab{}.
\newblock \showarticletitle{{DASA:} dissatisfaction-oriented advertising based
  on sentiment analysis}.
\newblock \bibinfo{journal}{\emph{Expert Systems with Applications}}
  \bibinfo{volume}{37}, \bibinfo{number}{9} (\bibinfo{year}{2010}),
  \bibinfo{pages}{6182--6191}.
\newblock


\bibitem[\protect\citeauthoryear{Saha, Chan, de~Barbaro, Abowd, and
  Choudhury}{Saha et~al\mbox{.}}{2017}]%
        {SahaCBAC17}
\bibfield{author}{\bibinfo{person}{Koustuv Saha}, \bibinfo{person}{Larry Chan},
  \bibinfo{person}{Kaya de Barbaro}, \bibinfo{person}{Gregory~D. Abowd}, {and}
  \bibinfo{person}{Munmun~De Choudhury}.} \bibinfo{year}{2017}\natexlab{}.
\newblock \showarticletitle{Inferring mood instability on social media by
  leveraging ecological momentary assessments}.
\newblock \bibinfo{journal}{\emph{Proceedings of the ACM on Interactive,
  Mobile, Wearable and Ubiquitous Technologies, {IMWUT}}} \bibinfo{volume}{1},
  \bibinfo{number}{3} (\bibinfo{year}{2017}), \bibinfo{pages}{95:1--95:27}.
\newblock


\bibitem[\protect\citeauthoryear{Si, Mukherjee, Liu, Li, Li, and Deng}{Si
  et~al\mbox{.}}{2013}]%
        {aclSiMLLLD13}
\bibfield{author}{\bibinfo{person}{Jianfeng Si}, \bibinfo{person}{Arjun
  Mukherjee}, \bibinfo{person}{Bing Liu}, \bibinfo{person}{Qing Li},
  \bibinfo{person}{Huayi Li}, {and} \bibinfo{person}{Xiaotie Deng}.}
  \bibinfo{year}{2013}\natexlab{}.
\newblock \showarticletitle{Exploiting topic based Twitter sentiment for stock
  prediction}. In \bibinfo{booktitle}{\emph{Proceedings of the 51st Annual
  Meeting of the Association for Computational Linguistics, {ACL} 2013}}.
  \bibinfo{pages}{24--29}.
\newblock


\bibitem[\protect\citeauthoryear{Sun, Guo, and Zhu}{Sun et~al\mbox{.}}{2019}]%
        {wwwSunGZ19}
\bibfield{author}{\bibinfo{person}{Lihua Sun}, \bibinfo{person}{Junpeng Guo},
  {and} \bibinfo{person}{Yanlin Zhu}.} \bibinfo{year}{2019}\natexlab{}.
\newblock \showarticletitle{Applying uncertainty theory into the restaurant
  recommender system based on sentiment analysis of online Chinese reviews}.
\newblock \bibinfo{journal}{\emph{World Wide Web}} \bibinfo{volume}{22},
  \bibinfo{number}{1} (\bibinfo{year}{2019}), \bibinfo{pages}{83--100}.
\newblock


\bibitem[\protect\citeauthoryear{Thelwall, Buckley, Paltoglou, Cai, and
  Kappas}{Thelwall et~al\mbox{.}}{2010}]%
        {ThelwallBPCK10}
\bibfield{author}{\bibinfo{person}{Mike Thelwall}, \bibinfo{person}{Kevan
  Buckley}, \bibinfo{person}{Georgios Paltoglou}, \bibinfo{person}{Di Cai},
  {and} \bibinfo{person}{Arvid Kappas}.} \bibinfo{year}{2010}\natexlab{}.
\newblock \showarticletitle{Sentiment in short strength detection informal
  text}.
\newblock \bibinfo{journal}{\emph{{JASIST}}} \bibinfo{volume}{61},
  \bibinfo{number}{12} (\bibinfo{year}{2010}), \bibinfo{pages}{2544--2558}.
\newblock


\bibitem[\protect\citeauthoryear{Wu, Zhang, Yuan, Wu, Huang, and Yan}{Wu
  et~al\mbox{.}}{2017}]%
        {WuZYWHY17}
\bibfield{author}{\bibinfo{person}{Fangzhao Wu}, \bibinfo{person}{Jia Zhang},
  \bibinfo{person}{Zhigang Yuan}, \bibinfo{person}{Sixing Wu},
  \bibinfo{person}{Yongfeng Huang}, {and} \bibinfo{person}{Jun Yan}.}
  \bibinfo{year}{2017}\natexlab{}.
\newblock \showarticletitle{Sentence-level sentiment classification with weak
  supervision}. In \bibinfo{booktitle}{\emph{Proceedings of the 40th
  International {ACM} {SIGIR} Conference on Research and Development in
  Information Retrieval, SIGIR 2017}}. \bibinfo{pages}{973--976}.
\newblock


\bibitem[\protect\citeauthoryear{Xiao and Guo}{Xiao and Guo}{2013}]%
        {XiaoG13}
\bibfield{author}{\bibinfo{person}{Min Xiao} {and} \bibinfo{person}{Yuhong
  Guo}.} \bibinfo{year}{2013}\natexlab{}.
\newblock \showarticletitle{Semi-supervised representation learning for
  cross-lingual text classification}. In \bibinfo{booktitle}{\emph{Proceedings
  of the 2013 Conference on Empirical Methods in Natural Language Processing,
  {EMNLP} 2013}}. \bibinfo{pages}{1465--1475}.
\newblock


\bibitem[\protect\citeauthoryear{Yang, Adamic, Ackerman, Wen, and Lin}{Yang
  et~al\mbox{.}}{2012}]%
        {YangAAWL12}
\bibfield{author}{\bibinfo{person}{Jiang Yang}, \bibinfo{person}{Lada~A.
  Adamic}, \bibinfo{person}{Mark~S. Ackerman}, \bibinfo{person}{Zhen Wen},
  {and} \bibinfo{person}{Ching{-}Yung Lin}.} \bibinfo{year}{2012}\natexlab{}.
\newblock \showarticletitle{The way I talk to you: sentiment expression in an
  organizational context}. In \bibinfo{booktitle}{\emph{Proceedings of the 2012
  International Conference on Human Factors in Computing Systems, {CHI} 2012}}.
\newblock


\bibitem[\protect\citeauthoryear{Zhang, Wang, and Liu}{Zhang
  et~al\mbox{.}}{2018}]%
        {ZhangWL18}
\bibfield{author}{\bibinfo{person}{Lei Zhang}, \bibinfo{person}{Shuai Wang},
  {and} \bibinfo{person}{Bing Liu}.} \bibinfo{year}{2018}\natexlab{}.
\newblock \showarticletitle{Deep learning for sentiment analysis: a survey}.
\newblock \bibinfo{journal}{\emph{Wiley Interdiscip. Rev. Data Min. Knowl.
  Discov.}} \bibinfo{volume}{8}, \bibinfo{number}{4} (\bibinfo{year}{2018}).
\newblock


\bibitem[\protect\citeauthoryear{Zhou, Pan, Tsang, and Ho}{Zhou
  et~al\mbox{.}}{2016a}]%
        {ZhouPTH16}
\bibfield{author}{\bibinfo{person}{Joey~Tianyi Zhou},
  \bibinfo{person}{Sinno~Jialin Pan}, \bibinfo{person}{Ivor~W. Tsang}, {and}
  \bibinfo{person}{Shen{-}Shyang Ho}.} \bibinfo{year}{2016}\natexlab{a}.
\newblock \showarticletitle{Transfer learning for cross-language text
  categorization through active correspondences construction}. In
  \bibinfo{booktitle}{\emph{Proceedings of the Thirtieth {AAAI} Conference on
  Artificial Intelligence, AAAI 2016}}. \bibinfo{pages}{2400--2406}.
\newblock


\bibitem[\protect\citeauthoryear{Zhou, Wan, and Xiao}{Zhou
  et~al\mbox{.}}{2016b}]%
        {ZhouWX16}
\bibfield{author}{\bibinfo{person}{Xinjie Zhou}, \bibinfo{person}{Xiaojun Wan},
  {and} \bibinfo{person}{Jianguo Xiao}.} \bibinfo{year}{2016}\natexlab{b}.
\newblock \showarticletitle{Cross-lingual sentiment classification with
  bilingual document representation learning}. In
  \bibinfo{booktitle}{\emph{Proceedings of the 54th Annual Meeting of the
  Association for Computational Linguistics, {ACL} 2016}}.
  \bibinfo{pages}{1403--1412}.
\newblock


\end{thebibliography}

\end{document}